\documentclass[%
 reprint,
 amsmath,amssymb,
 aps,
]{revtex4-2}

\usepackage{amsmath}
\usepackage[dvipsnames]{xcolor}
\usepackage{ragged2e}
\usepackage{float}
\usepackage{caption}
\usepackage{subcaption}
\usepackage{graphicx}
\usepackage{dcolumn}
\usepackage{bm}
\usepackage{lipsum, color}

\begin{document}

\preprint{}

\title{Quantifying Reconnection and it's Dynamical Role in 2D Magnetic Rayleigh-Taylor Turbulence}
\author{Manohar Teja Kalluri}
\email{manohartejakalluri25@gmail.com}
\author{Andrew Hillier}
\author{Ben Snow}
\affiliation{Department of Mathematics and Statistics, University of Exeter, Exeter, United Kingdom}

\date{\today}

\begin{abstract}
Magnetic Rayleigh-Taylor instability (MRTI) governs material transport and mixing in astrophysical and laboratory plasmas under the influence of gravity and magnetic fields. While magnetic reconnection is known to occur during MRTI evolution, its role in the evolution and energy dynamics remains poorly understood. Here, we present a comprehensive analysis of the role of reconnection in the two-dimensional MRTI dynamics, using high-resolution simulations. We establish that reconnection, through facilitating plume merger, relieving magnetic tension, and enabling continued instability growth, forms an essential component for the long-term instability evolution. To quantify the role of reconnection in energy dynamics, we develop a robust automated reconnection detection algorithm and perform a statistical analysis across a range of magnetic field strengths. We find that reconnection accounts for up to $80\%$ of the magnetic-to-kinetic energy transfer in the weak magnetic field regime, while contributing minimally ($\approx 3\%$) to magnetic energy dissipation. Our results establish magnetic reconnection as a critical mechanism that regulates large-scale MRTI dynamics, with implications for astrophysical plasmas and turbulent mixing in magnetized flows.
\end{abstract}

\maketitle


\section{Introduction} \label{intro}

\vspace{-10pt} The configuration of a high-density plasma being supported by a low-density plasma in the presence of gravitational and magnetic fields is ubiquitous in many laboratory \citep{Zhou2024} and astrophysical \citep{ZHOU2017_1, ZHOU2017_2} systems. Perturbations at the interface can lead to the interpenetration and mixing of the two plasmas, which is known as the magnetic Rayleigh-Taylor instability (MRTI). These structures of low-density plasma penetrating high-density plasma or vice versa are referred to as \textit{plumes}. The region where the plumes penetrate or the two plasmas mix is called the \textit{mixing layer}, which is typically a turbulent region \citep{Stone2007a}. The MRTI often plays a crucial role in the mixing or material transportation (through plumes) in numerous systems \citep{ZHOU2017_1, ZHOU2017_2}. For example, mixing of the corona and prominence materials in the solar prominence \citep{Hillier2016_solarReview}, transportation of the stellar material into the surrounding medium in supernova \citep{Gull1975, Bucciantini2004}, accretion of the disc material onto the central object in the accretion discs \citep{Kulkarni2008}. 

At the characteristic scales of the instability, plasma in most systems approximately obeys the ideal flux-freezing condition \citep{Dungey1953}, wherein magnetic field lines are tightly coupled to plasma flow. Consequently, perturbations deform both flow streamlines and magnetic field lines, with plasma motion dragging magnetic fields as the instability evolves. When turbulent motions bring oppositely directed magnetic field lines into close proximity, intense current sheets form due to steep magnetic gradients. In these regions, the current becomes strong enough to trigger non-ideal magnetohydrodynamic (MHD) behavior, breaking the flux-freezing condition and altering magnetic topology through magnetic reconnection \citep{priest_forbes_2000}.

Magnetic reconnection within the MRTI mixing layer has been illustrated by numerical simulations \citep{Hillier2012a, Popescu2021, Zhdankin_2023} and observations \citep{Hillier2012b}. Considering isolated reconnection events, these studies have shown that reconnection drives plasma downflows \citep{Hillier2012a, Hillier2012b}, and leads to plume mergers \citep{Popescu2021}.

Despite these advances, a key missing link in MRTI research is understanding the significance of magnetic reconnection in the large-scale evolution of the instability. Hydrodynamic Rayleigh-Taylor instability studies have established that plume mergers are essential to the instability’s nonlinear development \citep{Glimm1990, Li1995}. The role of magnetic reconnection in enabling plume mergers suggests that reconnection may be critical for the overall MRTI dynamics. This raises the question: \textit{Is magnetic reconnection an essential component of the long-term evolution of MRTI?}

Beyond morphology, magnetic reconnection is also expected to influence energy dynamics. In reconnection regions, magnetic energy converts into kinetic and thermal energies via the Lorentz force and current-sheet dissipation, making reconnection an important source of local plasma heating and flow acceleration \citep{Yamada2010}. However, in MRTI, gravitational potential energy (GPE) drives the instability by injecting kinetic energy, displacing fluid vertically, and broadening the mixing layer. As the instability evolves, magnetic field lines deform and often bundle around plumes \citep{Popescu2021}, forming dissipative current sheets. Thus, two energy sources coexist: GPE and magnetic reconnection. This motivates the question: \textit{Does magnetic reconnection significantly contribute to the energy dynamics of MRTI?}
 
Previous studies have largely focused on isolated reconnection events \citep{Hillier2012a, Popescu2021}, but the turbulent nature of the MRTI mixing layer \citep{Jun1995, Stone2007a, Carlyle2017, Kalluri_2024} allows the magnetic field lines to contact at multiple locations, leading to simultaneous reconnection events. Such simultaneous multiple reconnection events have been observed in homogeneous turbulence systems \citep{Servidio2010, Dong2022}. Unlike statistically steady homogeneous turbulence, the MRTI is dynamically evolving: the mixing layer grows approximately quadratically in time, and root-mean-square velocity increases roughly linearly \citep{Kalluri_2024}. Hence, the number and nature of reconnection events are expected to vary temporally.

Magnetic field strength is another critical factor influencing reconnection. Increasing the field strength suppresses small-scale perturbations and reduces turbulence intensity within the mixing layer \citep{Jun1995, Stone2007a, Carlyle2017, Kalluri_2024}, thereby modifying the turbulent dynamics of MRTI. Accordingly, the characteristics of reconnection are expected to vary with magnetic field strength. However, a systematic statistical characterization of this dependence remains lacking. This motivates the question: \textit{How do reconnection statistics and dynamics evolve with time and magnetic field strength?}

Addressing these questions, this paper aims to provide a comprehensive understanding of magnetic reconnection in MRTI. The qualitative role of reconnection in the large-scale instability evolution and energy dynamics is examined in Sections~\ref{evolution} and \ref{sec:energy_dynamics}, respectively. Section~\ref{Characteristics} presents a statistical analysis of reconnection events as functions of time and magnetic field strength. The quantitative contribution of reconnection to the energy dynamics is assessed in Section~\ref{sec:energydynamicsquant}.

\vspace{-10pt}
\section{Numerical methodology} \label{Methodology}

\begin{figure*}
    \centering
    \includegraphics[width = \textwidth]{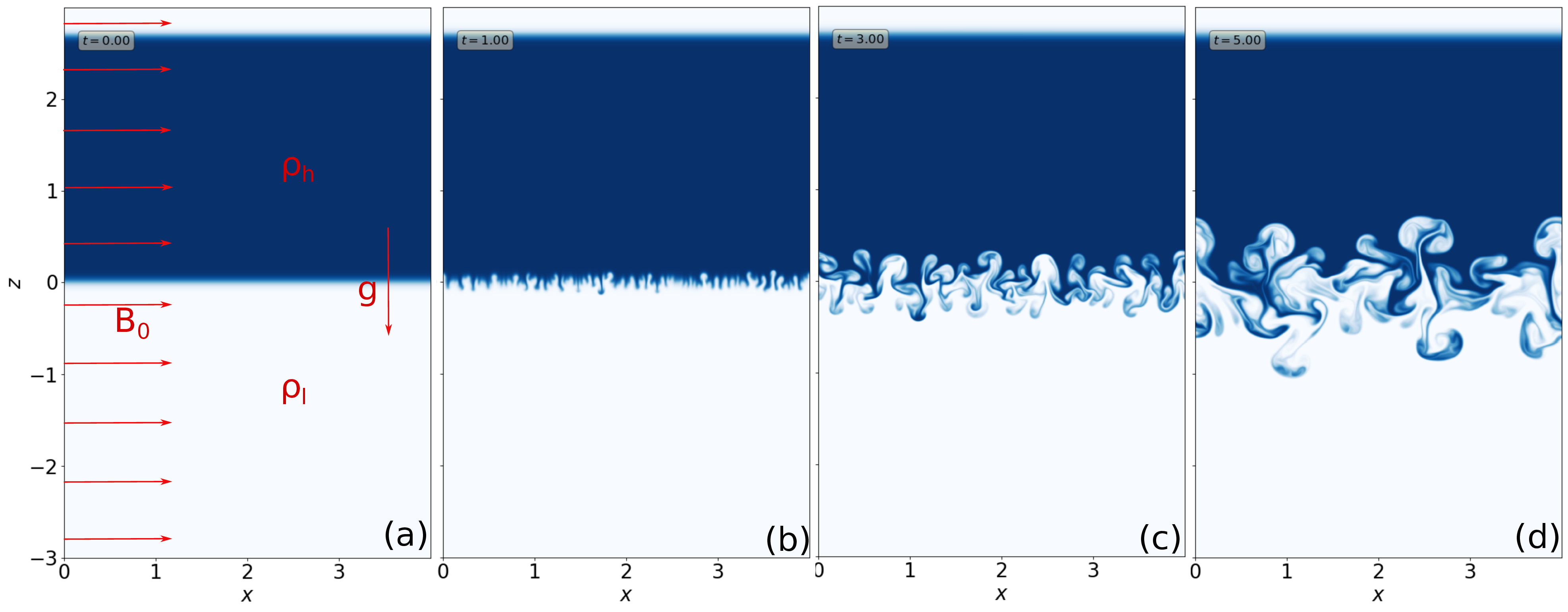}
    \caption{Figure showing the initial configuration (a), evolution (b, c, d) of magnetic Rayleigh-Taylor instability mixing layer through the density contours at different time instants $t {=} 1.0,$ $t {=} 3.0,$ $t {=} 5.0$ \textit{(from left to right)}. The snapshots correspond to $B_0 = 3\% B_c$ case.}
    \label{MRTI_config}
\end{figure*}

In the present study, MRTI is modelled numerically by superimposing a high density plasma $(\rho_h)$ over a low density plasma $(\rho_l)$ in the presence of an externally imposed uniform, unidirectional magnetic field $(\mathbf{B_0})$ and gravitational field $(\mathbf{g})$ as shown in figure \ref{MRTI_config}(a). The magnetic and gravitational fields are along the horizontal and vertical directions (parallel and perpendicular to the interface), respectively. The numerical modelling was performed using Dedalus \citep{Dedalus2020}, an open-source, parallelized computational framework to solve the partial differential equations using the psuedo-spectral method. We solve an approximation of the variable-density non-ideal MHD governing equations \citep{Kalluri_2024}
\begin{subequations}
    \begin{align}
        \partial_t \mathbf{u} {-} \nu \mathbf{\nabla}^2 \mathbf{u} = {-} (\mathbf{u} {\cdot} \mathbf{\nabla}) \mathbf{u} & {-} \frac{1}{\rho} \mathbf{\nabla} p {-} \frac{\delta \rho}{\rho}\mathbf{g} {-} \frac{1}{\rho} (\mathbf{B} {\cdot}  \mathbf{\nabla}) \mathbf{B}, \\
        \partial_t \mathbf{B} {-} \eta \mathbf{\nabla}^2 \mathbf{B} {-} c_p^2 \mathbf{\nabla}(\mathbf{\nabla} {\cdot} \mathbf{B}) & = (\mathbf{B} {\cdot} \mathbf{\nabla}) \mathbf{u} {-} (\mathbf{u} {\cdot} \mathbf{\nabla}) \mathbf{B}, \\
        \partial_t \rho {-} D \mathbf{\nabla}^2 \rho & = {-} (\mathbf{u} {\cdot} \mathbf{\nabla}) \rho,  \\
        \mathbf{\nabla} {\cdot}  \mathbf{u} & = 0.
    \end{align}
    \label{MHDeqns_1}
\end{subequations}
$\textbf{u}$, $\textbf{B}$, $\rho$ represent the velocity, magnetic field, and density, respectively. The bold letter is used to specify a vector quantity.  Note that $\nabla \cdot u = 0$ is the incompressible limit of the true velocity-divergence of the variable-density approximation \citep{livescu_ristorcelli_2008}. The solenoidal condition for magnetic field ($\mathbf{\nabla} \cdot \mathbf{B} = 0$) is ensured through divergence cleaning term ($c_p^2 \mathbf{\nabla}(\mathbf{\nabla} \cdot \mathbf{B})$) \citep{DEDNER_2002}. In the present study, the value of $c_p$ is set such that the solenoid condition is satisfied to machine precision throughout the simulation. We consider two miscible plasmas with equal and constant density diffusion coefficient ($D$) \citep{Briard_Gréa_Nguyen_2024}. The diffusion smooths sharp density gradients and aids mixing of plasma at the grid scale. $\nu, \eta$ are the coefficients of fluid viscosity and magnetic diffusion, respectively. $\nu, \eta,$ and $D$ are set to $10^{-4}$. We use Runge-Kutta scheme of fourth-order for time stepping, integrating the linear terms implicitly and the non-linear terms explicitly (see $\S$IX F of \cite{Dedalus2020}). Dedalus does not use an explicit linear pressure solver. The value of pressure is determined by the divergence-free velocity constraint.

Here, the MRTI is studied in 2D with periodic boundary conditions in both directions. A domain of length $L_x {\times} L_z {=} 4 {\times} 6$ units $(x:[0, L_x], z:[-L_z/2,L_z/2])$ with resolution of $2048 {\times} 3072$ is taken. The acceleration due to gravity is taken as 1. All the MRTI cases were run at a density ratio of 3. That is, all simulations have an Atwood number, defined as $\left( \frac{\rho_h - \rho_l}{\rho_h + \rho_l} \right)$, of 0.5.

The initial density profile is given by equation \ref{rho_profile} 
\begin{equation}
\begin{split}
    \rho = \rho_l {-} \frac{\Delta \rho}{2} \left[ \tanh{\left(\frac{z {-} 0.45 L_z}{0.05} \right)} {+} 1 \right] {+} \frac{\Delta \rho}{2} \left[ \tanh{\left(\frac{z}{0.05} \right)} {+} 1 \right]
    \label{rho_profile}
\end{split} 
\end{equation}
where $\Delta \rho {=} \rho_h {-} \rho_l$. The profile results in two interfaces, one at $z {=} 0$ and the other at $z {=} 0.45 L_z (z = 2.7)$ as shown in figure \ref{MRTI_config}(a). The transition between the two densities is made continuous using a hyperbolic tangent profile with a half width of $l {=} 0.05$ (see equation \ref{rho_profile}) spanning 25 grid points. This ensures that the interface is resolved sufficiently throughout the instability evolution despite the thinning of the interface. The above density profile is chosen to ensure periodicity in the $z$-direction.

The MRTI is excited by perturbing the lower interface (at $z = 0$) with a vertical velocity of the form,
\begin{equation}
    u_z {=} \sum^{128}_{i=1} a_i \sin \left( \frac{2 \pi k_i x}{L_x} {+} \phi_i \right) e^{{-}(z^2/0.01)}.
    \label{white_noise_2d}
\end{equation}
The perturbations decay in a Gaussian profile about the interface. The amplitude $(a)$, wave mode $(k)$, and phase $(\phi)$ of the perturbation are chosen between [-0.0125, 0.0125], [1, 128], and [0, $\pi$], respectively. The perturbation has a wide range of wave modes (as suggested by the previous studies \citep{ramaprabhu_2005, dalziel_1999, Dimonte2004, GLIMM2001}) along $x$. 

To test the variation of reconnection dynamics with magnetic field strength, we run MRTI simulations over a range of magnetic field strengths between $1\%$ and $25\% B_c$. Here, $B_c$ refers to the magnetic field strength at which the wave modes parallel to the magnetic field are completely suppressed. $B_c$ is given by \citep{Chandrasekhar1961}
\begin{equation}
    B_c = \sqrt{\frac{(\rho_h - \rho_l) g}{2 k}}.
    \label{Bc}
\end{equation}
Note that, $B_c$ is originally defined for a single-mode undular perturbation. For consistency, we choose the $B_c$ based on the smallest mode, $k/L_x = 1$. We non-dimensionalise the imposed magnetic field strength with $B_c$, as a way to characterise the imposed magnetic field strength.

\section{Reconnection and evolution of MRTI} \label{evolution}

In $\S$\ref{intro}, we posed the question of whether reconnection is necessary for the long-term evolution of MRTI, analogous to the role of plume mergers in hydrodynamic RTI. To investigate this, we simulate two scenarios: an idealized MRTI where reconnection does not occur ($\S$\ref{IdealMRTI}), and a non-ideal MRTI where reconnection occurs ($\S$\ref{section:pinching}). Comparing these two cases, we infer the necessity of reconnection.

\vspace{-10pt}
\subsection{The pseudo-ideal-MRTI case} \label{IdealMRTI}

\begin{figure*}
    \centering
    \includegraphics[width = \textwidth]{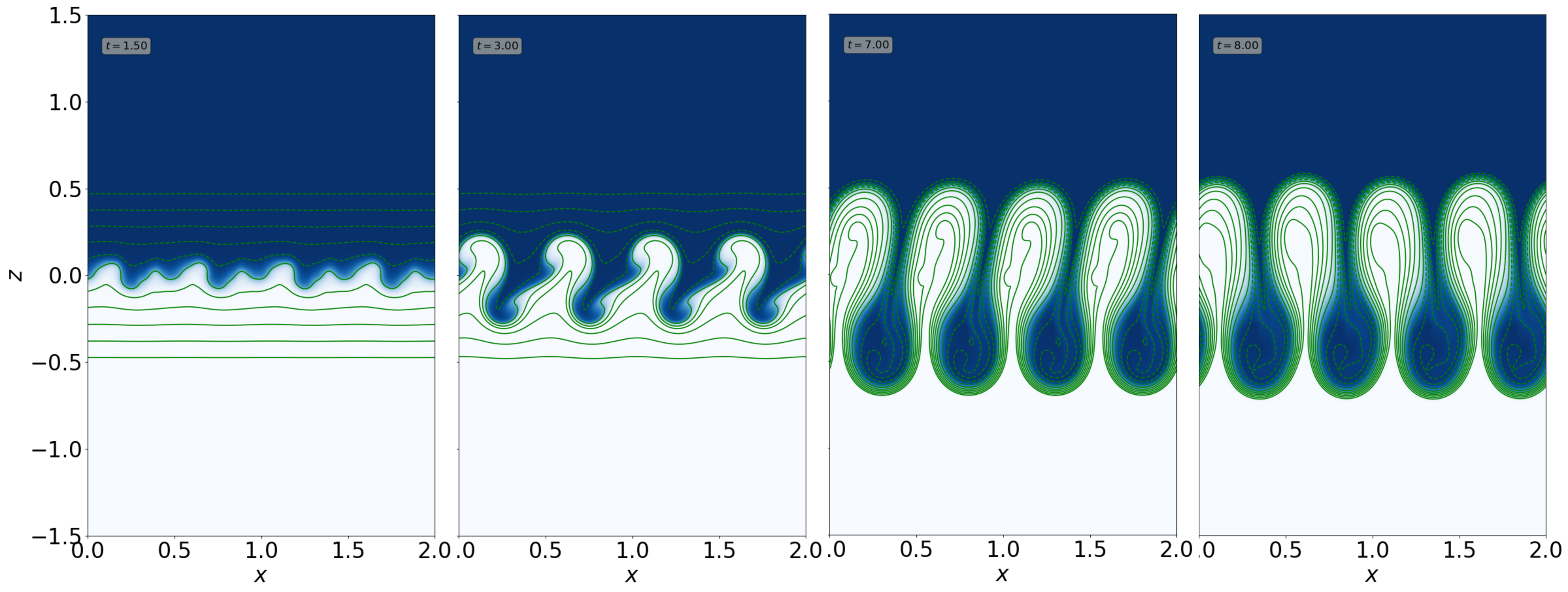}
    \caption{Instantaneous density contours showing the evolution of (analogous) ideal MRTI at different time instants: (a) $t = 1.50$, (b) $t = 3.0$, (c) $t = 7.0$, (d) $t = 8.0$. The white and blue regions represent the low-density ($\rho_l = 1$) and high-density ($\rho_h = 3$) plasma, respectively. The red lines represents the 11 field lines of different magnetic vector potentials ($\mathcal{A}$, calculated from equation \ref{vec_pot})  ranged between [-0.05, 0.05]. $\mathcal{A} < 0$ are dashed and $\mathcal{A} \geq 0$ are solid.}
    \label{norec}
\end{figure*}

\begin{figure}
    \centering
    \includegraphics[width = 0.7\linewidth]{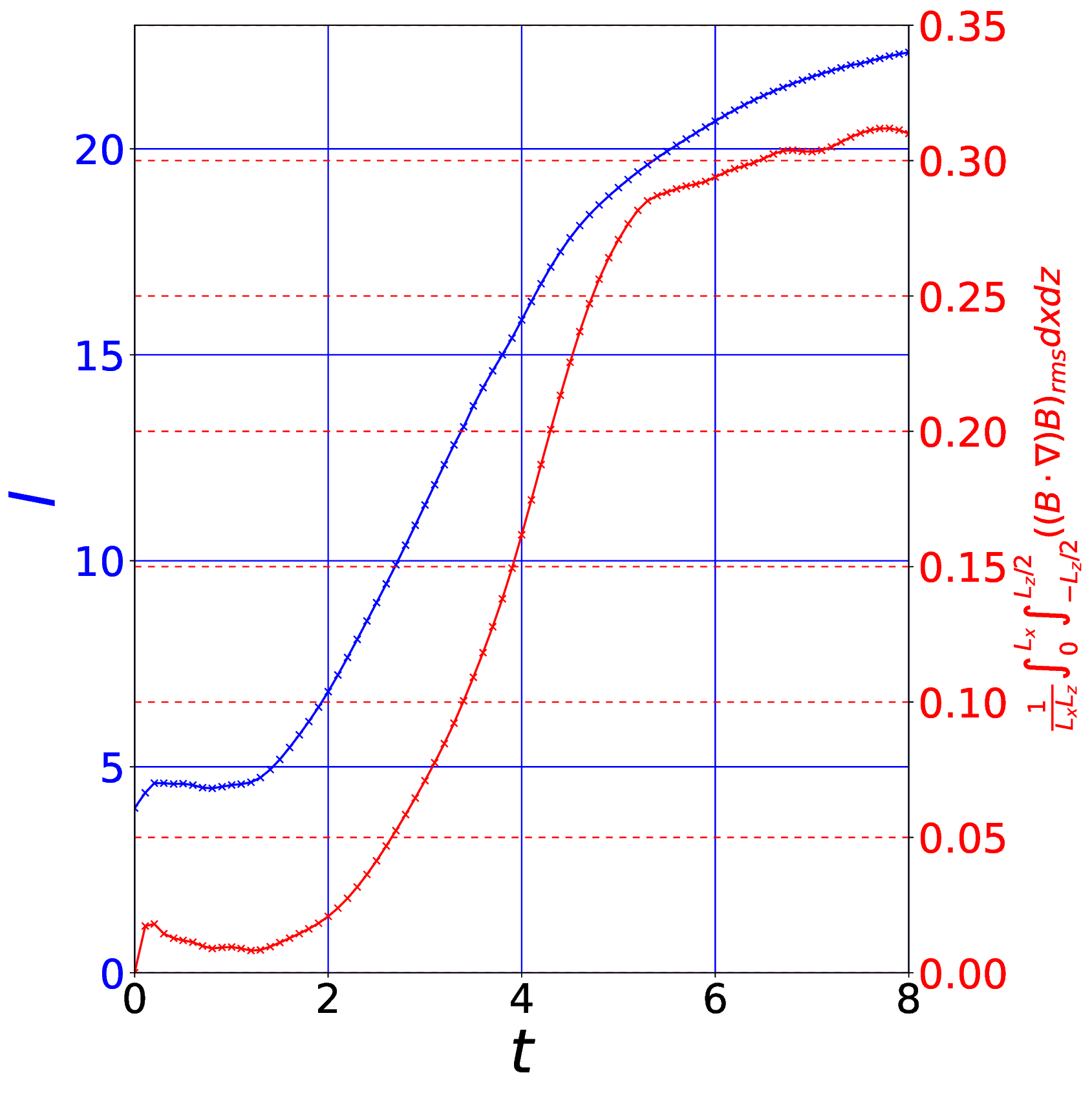}
    \caption{Temporal variation of: length of the magnetic field line initially at $z=0$ (left axis) and magnetic tension (right axis) for the ideal-MRTI. }
    \label{l_bb}
\end{figure}
\begin{figure}
    \centering
    \includegraphics[width = 0.75\linewidth]{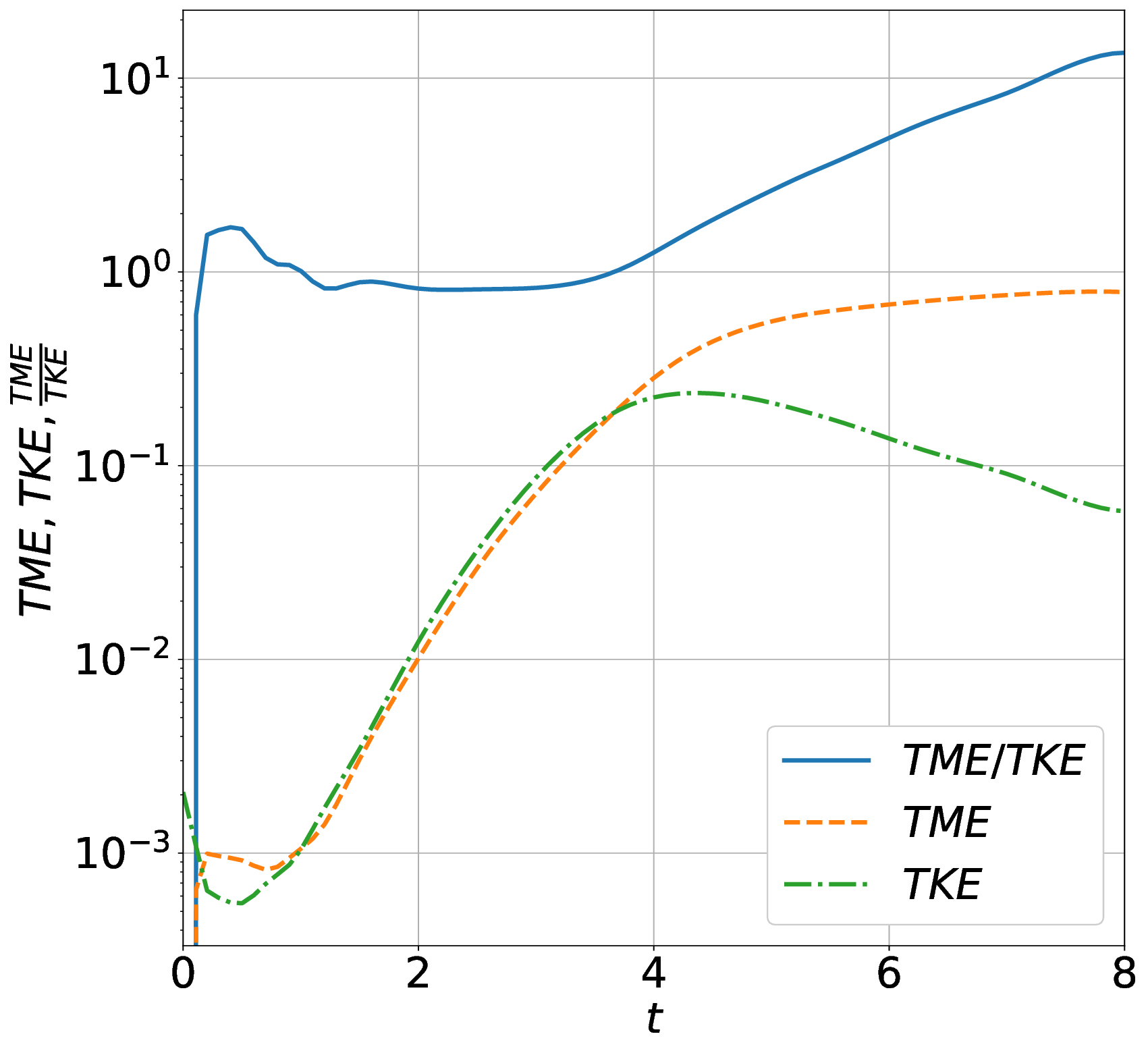}
    \caption{Temporal variation of turbulent magnetic energy (TME, $\textcolor{orange}{\textbf{- -}}$), turbulent kinetic energy (TKE, $\textcolor{ForestGreen}{\textbf{-.}}$), and their ratio ($\textcolor{blue}{\textbf{---}}$) for the ideal-MRTI.}
    \label{energy}
\end{figure}
\vspace{-10pt} A truly ideal MRTI, where $D = \nu = \eta = 0$, is numerically infeasible in Dedalus due to inherent stability constraints. Alternately, this can be simulated by using low $D, \nu,$ and $ \eta$ such that only numerical diffusion drives the non-ideal physics, while ensuring $D, \nu,$ and $ \eta$ are large enough to avoid numerical artifacts like ringing. However, this still does not prevent reconnection, the key physics of interest in the current study, since the reconnection occurs due to the numerical diffusion in the codes. In the context of the present study, understanding the role of reconnection, the key aspect is not the magnitude of the diffusion coefficients, but avoiding the reconnection. Therefore, we simulate an analogous ideal-MRTI case where reconnection is prevented by using a carefully designed perturbation (equation \ref{pert_norec}) and a strong magnetic field. The strong magnetic field ($B_0 = 15\% B_c$ in the present study) suppresses small wavelength perturbations \citep{Jun1995} ($k > k_c \approx 44$ in the present study). As such, no reconnection occurs at the early stages of the instability. The reconnection at the later stages of instability is due to integral length scale plumes, i.e., large wavelength perturbations, which are avoided by limiting the smallest wave mode to 8 in our perturbation. Hence, we choose a  perturbation of the form
\begin{equation}
    u_z {=} \sum^{100}_{k_i=8} a_i \sin \left(\frac{2 \pi k_i x}{L_x} + \phi_i \right) e^{{-}(z^2/0.01)},
    \label{pert_norec}
\end{equation}
where $a_i \in [-0.05, 0.05]$, $k_i \in [8, 100]$, and $\phi_i \in [0, \pi]$. The values of $\nu, \eta, D, L_x, L_z, N_x, N_z$ are same as mentioned in $\S$\ref{Methodology}.

\begin{figure*}
    \centering
    \includegraphics[width = \textwidth]{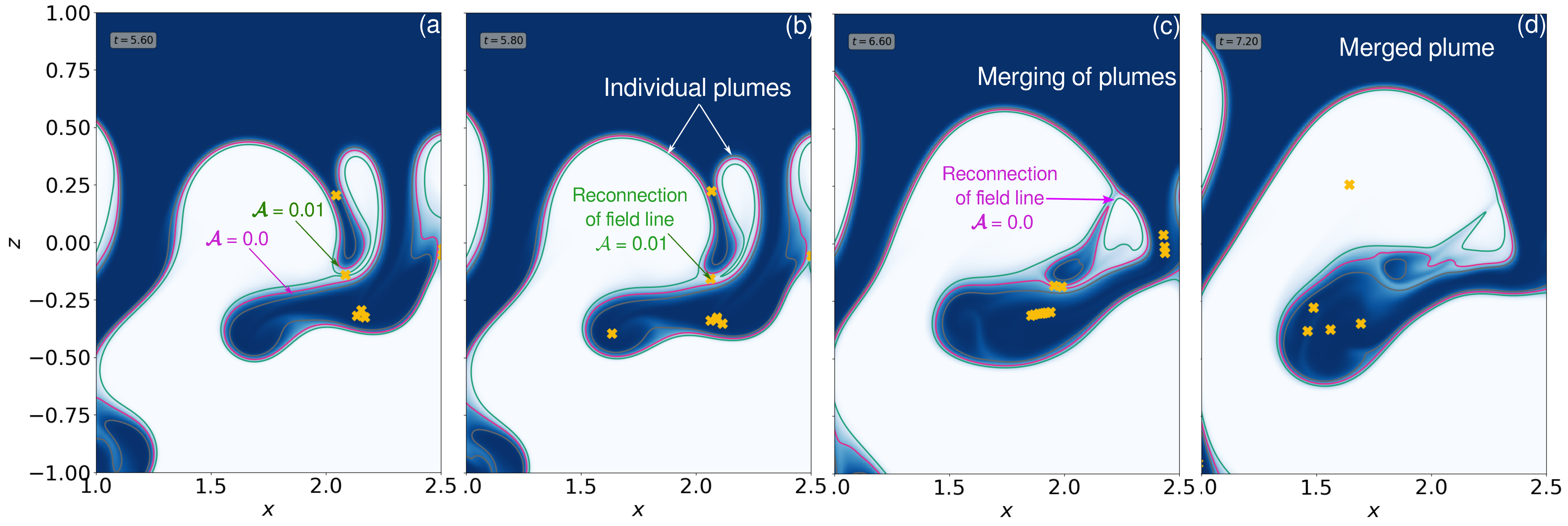}
    \caption{A snippet of the mixing layer of non-ideal-MRTI showing the merger of plumes over time: (a) $t = 5.6,$ (b) $t = 5.8,$ (c) $t = 6.6$, (d) $t = 7.2$. The change in magnetic field line topology during the magnetic reconnection process is shown. The yellow $\mathbf{\times}$ mark denotes the reconnection event. The black lines are magnetic field lines, calculated from equation \ref{vec_pot}.}
    \label{merge}
\end{figure*}
The evolution of the ideal-analogue-MRTI (along with the magnetic field lines) is shown in figure \ref{norec}. The magnetic field lines correspond to different values of magnetic potential ($\mathcal{A}$), given by $B(x, z) = \nabla \times \mathcal{A}$, a one-dimensional vector in a 2D problem. We calculate the $\mathcal{A}$ in the post-processing stage as
\begin{equation}
    \mathcal{A} = \int_{0}^{x} B_z (X, 0) \mathrm{d}X - \int_{-L_z/2}^{Z} B_x (x, Z) \mathrm{d}Z + C,
    \label{vec_pot}
\end{equation}
where $C = \int_{0}^{x} B_z (X, 0) \mathrm{d}X - \int_{-L_z/2}^{0} B_x (x, Z) \mathrm{d}Z$ is the gauge constant, chosen from a reference point ($(0, -L_z/2)$ in the current study) such that $\mathcal{A} = 0$ at the interface ($z = 0$) at time $t = 0$. Due to the change in value of $B_z (0, -L_z/2)$, we see an error in the gauge constant by around $2\%$ at the end of the simulation, where the mixing layer reaches the bottom boundary of the domain. The magnetic potential throughout the study is calculated as described above. As the instability grows, the magnetic field lines elongate as shown qualitatively and quantitatively in figures \ref{norec}, \ref{l_bb}, respectively. The magnetic field line initially at $z = 0$ deforms early in the instability, providing a representative measure of magnetic tension buildup. We track its temporal length, which correlates with the magnetic tension (figure \ref{l_bb}, right axis). As the magnetic field lines elongate and saturate, the magnetic tension also increases and saturates correspondingly. This demonstrates that the length of the magnetic field line is a good tool to estimate the magnetic tension in the system. Note that the correlation between field line length and tension depends on field line location; central lines (e.g., at $z=0$) offer the most representative measure as they evolve from the time $t=0$.

During the instability growth, the turbulent magnetic energy (TME) and the turbulent kinetic energy (TKE) increase equally (see figure \ref{energy}, $t \in [{\approx} 1, {\approx} 3.75]$). The TME and TKE are defined as 
\begin{equation}
    \text{TME} = \int_{0}^{L_x} \int_{-L_z/2}^{L_z/2} \left( \frac{1}{2} B^2 - \frac{1}{2} B_0^2 \right) \mathrm{d}x \mathrm{d}z,
\end{equation}
\begin{equation}
    \text{TKE} = \int_{0}^{L_x} \int_{-L_z/2}^{L_z/2} \left( \frac{1}{2} \rho u^2 \right) \mathrm{d}x \mathrm{d}z.    
\end{equation}
As the plumes evolve to integral length-scales, the energy required to bend the magnetic field lines increases, resulting in an unequal distribution of GPE between TKE and TME. A greater proportion of GPE is converted to magnetic energy. Consequently, the proportion of GPE converted to TKE reduces. This forms a positive feedback loop that leads to a continuous increase in TME (and hence a continued decrease in TKE), see figure \ref{energy}, $t \gtrapprox 3.75$. 

Over time, the magnetic tension becomes strong enough to balance the magnetic buoyancy force and arrest the further growth of instability, as shown in figures \ref{norec}(c) and (d) where the mixing layer height is approximately the same at $t \approx 7, 8$. At this stage, the plumes push against each other and oscillate left and right (figures \ref{norec}(c), (d)). 

To summarize, we showed that the length of the magnetic field line represents the magnetic tension in the system. As the instability evolves, the length of the magnetic field lines and with that the magnetic tension increases. If the tension is not relieved, over time the magnetic tension force becomes strong enough to counter the gravity and arrest further growth of instability. Thus the magnetic fields saturate the instability growth over time. 

\vspace{-10pt}
\subsection{The non-ideal-MRTI} \label{section:pinching}

\vspace{-10pt} To understand how reconnection influences the evolution, we will now consider the case of non-ideal-MRTI. As mentioned before, we change the perturbation to replicate the physics of ideal and non-ideal MRTI physics. Hence, we keep all the parameters the same between $\S$\ref{IdealMRTI} and $\S$\ref{section:pinching} except for the velocity perturbation. The numerical methodology of the non-ideal-MRTI simulation, initial perturbation, $\nu, \eta, D, L_x, L_z, N_x, N_z$ is as explained in $\S$\ref{Methodology}. For comparison with the ideal-MRTI, we use the same field strength as the ideal-MRTI, $B_0 {=} 15\%B_c$. Due to the small $\eta (= 10^{-4})$, the field lines evolve approximately with the flow.

\begin{figure}
    \centering
    \includegraphics[width=0.7\linewidth]{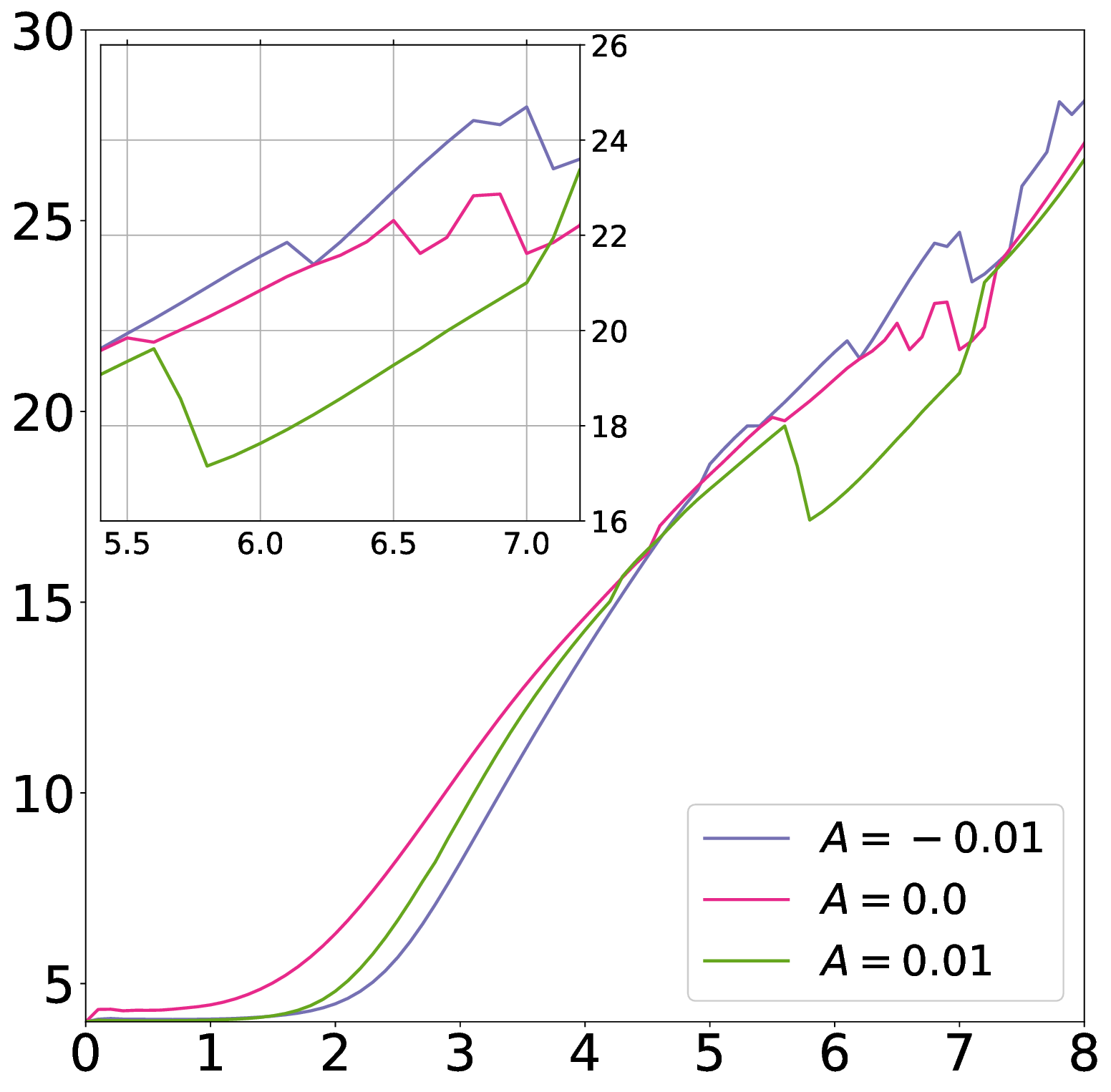}
    \caption{Figure showing the length of the magnetic field line with time. The inset figure shows the length during the time interval $t \in [5.4, 7.2]$}
    \label{length}
\end{figure}

\begin{figure}
    \centering
    \includegraphics[width=0.7\linewidth]{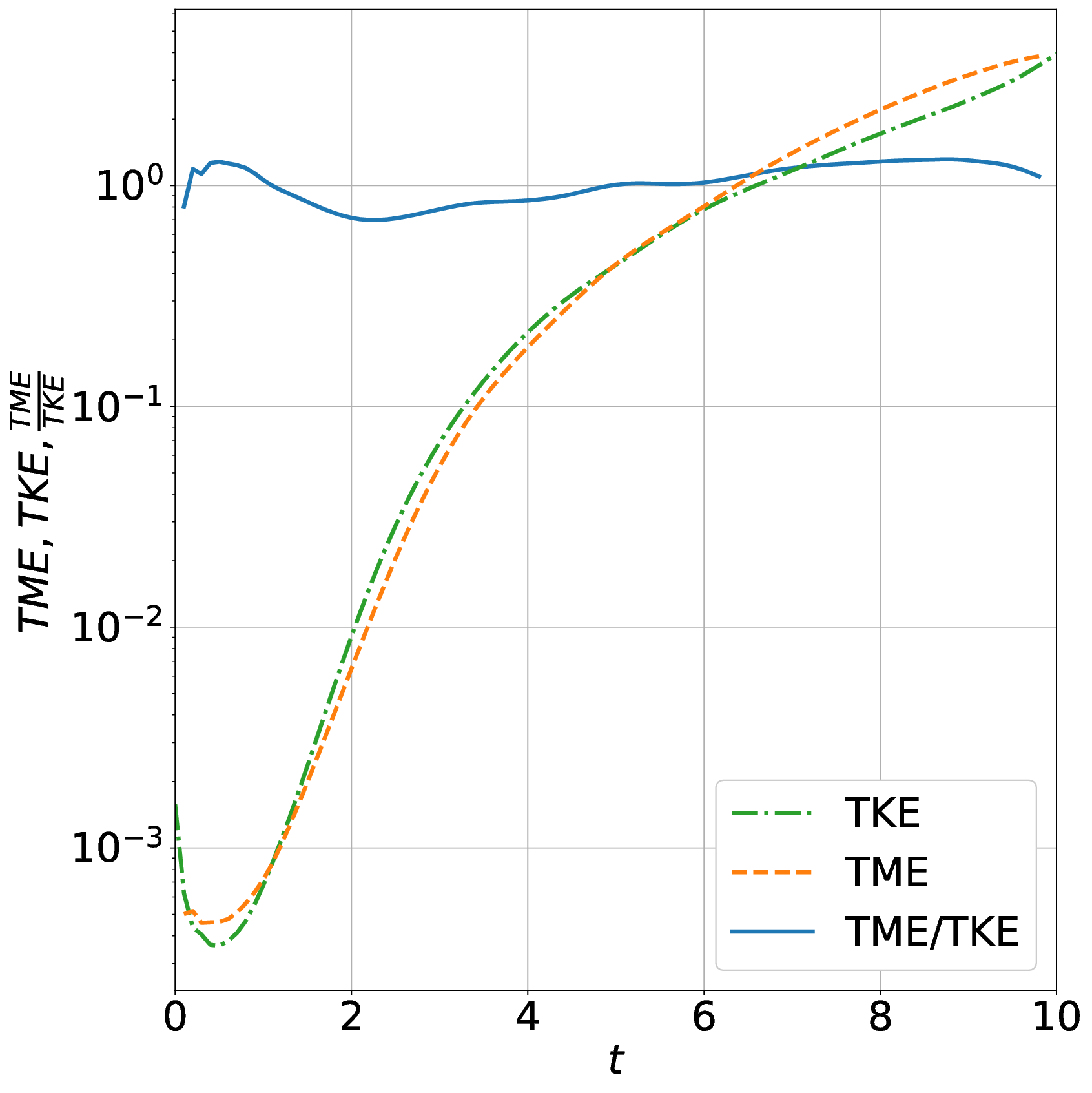}
    \caption{Temporal variation of turbulent magnetic energy (TME, $\textcolor{orange}{\textbf{- -}}$), turbulent kinetic energy (TKE, $\textcolor{ForestGreen}{\textbf{-.}}$), and their ratio ($\textcolor{blue}{\textbf{---}}$) for the non-ideal-MRTI.}
    \label{TKE_TME}
\end{figure}

Similar to the ideal-MRTI case, the strong magnetic field suppresses most small-scale wave modes, and the mixing layer evolves as laminar plumes. However, the non-linear interaction of large-scale plumes results in the contact of magnetic field lines around them. Due to the non-zero magnetic diffusivity, the magnetic field lines reconnect. Such a reconnection is shown in figures \ref{merge}(a), (b) using a sample field line ($\mathcal{A} = 0.01$). The magnetic reconnection leads to shorter magnetic field lines. The reduction in the length of the field line at $t \approx 5.8$ (when the reconnection occurs) is shown in the inset figure of \ref{length}. The shortening of field line length meant that the magnetic tension is relieved (c.f. $\S$\ref{IdealMRTI}). Thus, the magnetic reconnection prevents the domination of TME over TKE. This is illustrated in figure \ref{TKE_TME} where the system is maintained at approximate energy equipartition throughout the simulation (unlike the ideal-MRTI).

The domination of non-ideal processes at these reconnecting locations results in the transportation of plasma across the reconnecting field lines, allowing the two plumes to coalesce (see figure \ref{merge}(c)) towards forming a larger plume. The merger process over time is shown in figure \ref{merge}. 

Now, consider the forces acting on the plumes. The plume is under the influence of two forces: $i)$ buoyancy force ($F_b = (\rho_h - \rho_l) V_{plume} g$), which tries to displace the plume upward, $ii)$ magnetic tension, which opposes the buoyancy force. The length of the field line $\mathcal{A} = 0.01$ decreases from $\approx 20$ to $\approx 17$ due to reconnection at $t \approx 5.6$ (see figure \ref{length}). Similarly, the reconnection of field line $\mathcal{A} = 0.0$ at $t \approx 6.6$, also shortens its length. On the other hand, the buoyancy force $F_b$ is increased due to a larger plume volume. Arresting the growth of the merged plume demands a larger magnetic tension. However, the magnetic tension reduces due to reconnection, making it insufficient to counter the increased buoyancy force due to plume merger. Thus, there is a magnetic tension deficit that leads to further growth of the plume and the instability.

\vspace{-10pt}
\subsection{Section summary}

\vspace{-10pt} 
In summary, the ideal MRTI evolves until increasing magnetic tension halts further growth. In contrast, in non-ideal MRTI, reconnection relieves magnetic tension by changing magnetic field topology and reducing the field line length. Further, reconnection allows the merger of the interacting plumes, forming a bigger plume that experiences a greater buoyancy force. The merged plume requires higher magnetic tension to be arrested, but this tension is diminished by reconnection. Thus, magnetic reconnection enables sustained instability growth, demonstrating its critical role in MRTI evolution.

\section{Energy dynamics due to reconnection: Qualitative} \label{sec:energy_dynamics}
\begin{figure*}
    \centering
    \includegraphics[width = \textwidth]{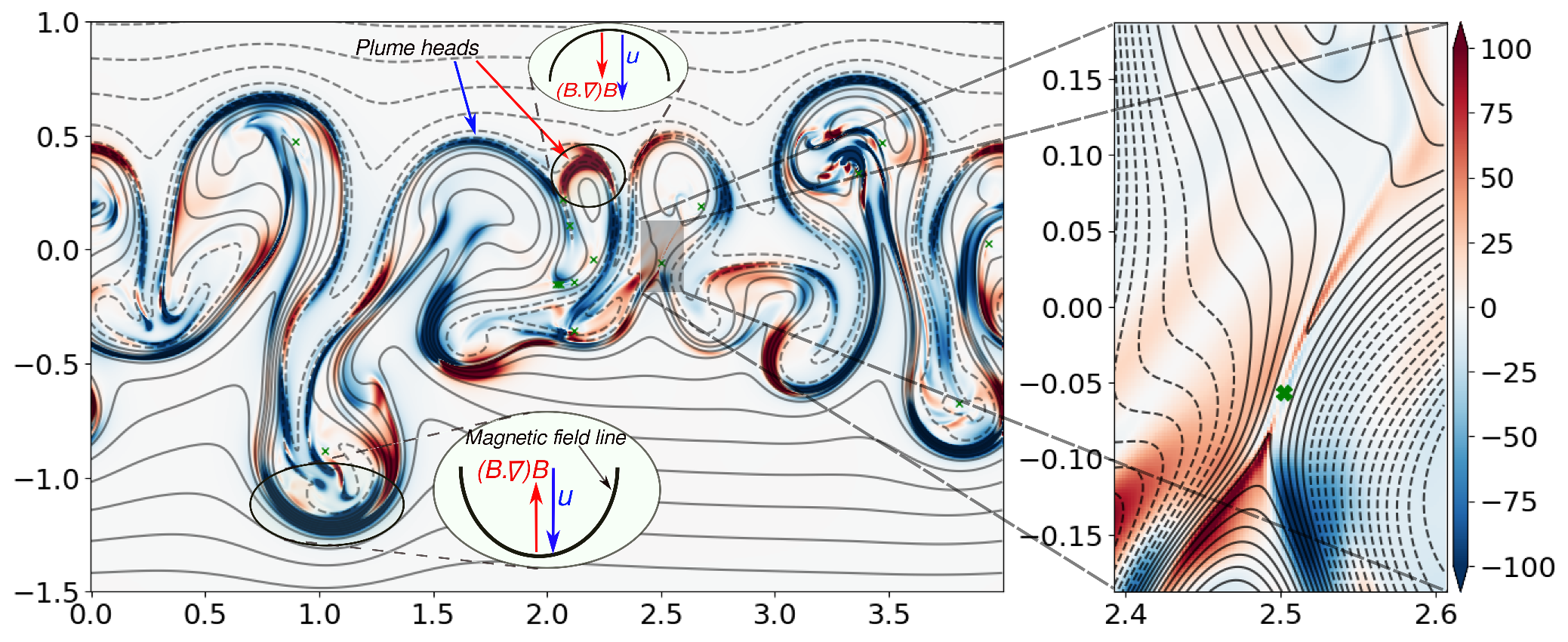}
    \caption{Instantaneous spatial contour of $u {\cdot} (B {\cdot} \nabla) B$ normalized with $\eta j_{rms}^2$: (left) full domain; (right) zoomed in around a reconnection point. \textcolor{ForestGreen}{$\times$} represent the reconnection points.}
    \label{ubdb_contour}
\end{figure*}

\vspace{-10pt} In $\S$\ref{intro}, we discussed how magnetic reconnection contributes to the energy dynamics of MRTI. To evaluate its significance, we first need to understand how reconnection energy dynamics manifest in the system. In this section, we qualitatively illustrate the conversion of magnetic energy into kinetic and thermal energy during reconnection events. The quantitative evaluation of these processes, which requires a statistical approach, is deferred to $\S$\ref{sec:energydynamicsquant}.

\vspace{-10pt}
\subsection{Energy interplay between kinetic and magnetic energy} \label{energyinterplay}

\vspace{-10pt} To understand the interplay between kinetic and magnetic energy, we consider the term $\mathbf{u} \cdot (\mathbf{B} \cdot \nabla) \mathbf{B}$, which arises in the turbulent kinetic energy (TKE) equation. The term is a product of flow velocity ($\mathbf{u}$) and the Lorentz term ($(\mathbf{B} \cdot \nabla) \mathbf{B}$. Here, $\mathbf{B}$ denotes the total magnetic field, including both the background and perturbations. This term quantifies the work done by magnetic tension on the fluid: it is positive when magnetic energy is transferred to the flow, and negative when kinetic energy is expended against magnetic tension.

Buoyancy forces drive the plumes to rise or fall, deforming the magnetic field and generating magnetic tension that resists this motion. A schematic of the opposing flow and magnetic tension is provided in figure \ref{ubdb_contour} (bottom inset). Due to the opposing nature of flow and magnetic tension, the term $\mathbf{u} \cdot (\mathbf{B} \cdot \nabla) \mathbf{B}$ would be negative (see the plume heads in figure \ref{ubdb_contour}). However, some plume heads exhibit positive values of $\mathbf{u} \cdot (\mathbf{B} \cdot \nabla) \mathbf{B}$ (one such is marked with a red arrow), indicating that magnetic tension and plume motion are aligned. This alignment likely results from interactions with neighboring plumes. For instance, the larger plume (blue arrow) exerts a downward force on the adjacent smaller plume (red arrow). The figure \ref{merge}(a) denotes the approximate time instant as the figure \ref{ubdb_contour}. From figure \ref{merge}, we find that the small plume which is initially at $(x, z) \approx (2.1, 0.25)$ has a downward motion due to the adjacent large plume. The downward motion of the small plume due to the large plume can be seen from figure \ref{merge}. Thus, the direction of magnetic tension and the net flow of the smaller plume are the same, making the quantity $\mathbf{u} \cdot (\mathbf{B} \cdot \nabla) \mathbf{B}$ positive. A schematic of the flow and magnetic tension is shown in figure \ref{ubdb_contour} as an inset at the top. 

To demonstrate the influence of reconnection on the energy dynamics, we zoom into a chosen reconnection point (see figure \ref{ubdb_contour}(right)). Near the reconnection site, we observe outflows characterized by positive $\mathbf{u} \cdot (\mathbf{B} \cdot \nabla) \mathbf{B}$, indicating localized conversion of magnetic energy into kinetic energy. The influence of reconnection can be seen to extend far beyond the reconnection point. The injection of kinetic energy into the system improves the fluid mixing locally.

For such distinct dynamics to appear, the reconnecting current sheets should be strong, and the turbulence surrounding the reconnection point should be minimal. But the majority of the reconnection events occur in turbulent regions and due to weak current sheets (evidenced in $\S$\ref{Characteristics}). Hence, such distinct dynamics are uncommon. 

\vspace{-10pt} \subsection{Energy dissipation:} \label{dissipation}
\begin{figure*}
    \centering
    \includegraphics[width = \textwidth]{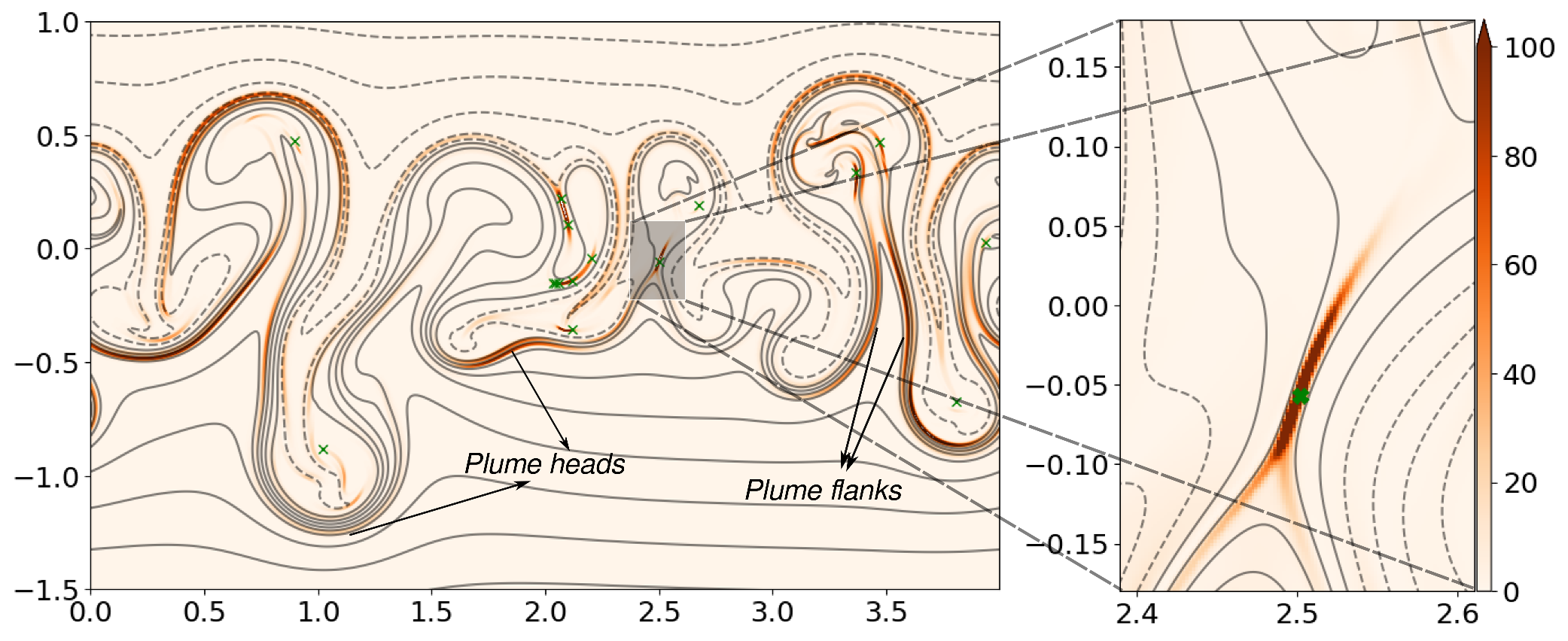}
    \caption{Instantaneous spatial contour of magnetic dissipation normalised with $\eta j_{rms}^2$: (left) full domain; (right) zoomed in around a reconnection point. \textcolor{ForestGreen}{$\times$} represent the reconnection points.}
    \label{mag_diss_contour}
\end{figure*}

\vspace{-10pt} In addition to kinetic energy conversion, reconnection also leads to the dissipation of magnetic energy. Consequently, reconnection regions are also the sites of significant dissipation. In a homogeneous turbulent system, the reconnection sites are major contributors to TME dissipation \citep{Servidio2010, Dong2022}. However, in MRTI, current sheets also arise from bending of magnetic field line around plume heads. This makes the relative contribution of reconnection to global magnetic energy dissipation less clear.

To understand this, we plot the spatial contours of instantaneous magnetic energy dissipation (figure \ref{mag_diss_contour}). The magnetic energy dissipation is calculated as
\begin{equation}
    D_{TME} = \int_0^t \int_0^{L_x} \int_{-L_z/2}^{L_z/2} \eta (\partial_j b_i)^2 \mathrm{d}x \mathrm{d}z \mathrm{d}t.
\end{equation}
Upon zooming into the neighborhood of a reconnection point (marked by $\textcolor{ForestGreen}{\times}$), we observe locally intense dissipation (see figure \ref{mag_diss_contour}(right)). However, significant dissipation is also observed in non-reconnecting current sheets, particularly near plume heads and flanks, where magnetic fields are highly concentrated due to the opposing nature of flow and magnetic tension. The concentration of these field lines leads to strong current sheets that dissipate magnetic energy significantly. 

Since reconnection sites are relatively small in spatial extent compared to other current sheets, their contribution to global magnetic energy dissipation may be limited. A quantitative analysis requires detecting the reconnection regions, which will be discussed in $\S$\ref{section:detection}. Following this, a quantitative analysis of energy dissipation due to reconnection will be made in $\S$\ref{tmedissquant}.

\vspace{-10pt}
\subsection{Section summary}
\vspace{-10pt} In summary, we qualitatively examined how magnetic reconnection contributes to energy dynamics in MRTI. We illustrated the conversion of magnetic energy into kinetic energy via reconnection-driven outflows, whose influence extends well beyond the reconnection site. However, such clear signatures are rare and typically suppressed by surrounding turbulence. Additionally, we highlighted that while reconnection does lead to magnetic energy dissipation, significant dissipation also occurs in non-reconnecting current sheets due to the natural flow–field interactions in MRTI. A more quantitative treatment follows in Sections~\ref{Characteristics} and \ref{sec:energydynamicsquant}.

\section{Statistics of MRTI reconnection} \label{Characteristics}
\vspace{-10pt} 
To this point, the role of reconnection was elucidated using a sample reconnection point. However, MRTI has simultaneous multiple reconnection events, which change quantitatively with time and magnetic field strength (see $\S$\ref{intro}). Towards a statistical analysis of the reconnection, we propose a reconnection detection algorithm in $\S$\ref{section:detection}. Using the algorithm, we explore the time evolution of the number of reconnection points in $\S$\ref{sec:number}, and characterize their strength in $\S$\ref{sec:characteristics}.  In particular, we focus on the fraction of strong reconnection current sheets, since they are expected to have a major influence on the energy dynamics of the system. 

\vspace{-10pt}
\subsection{Detection of reconnection points} \label{section:detection}
\vspace{-10pt} The first step towards quantifying the reconnection points is to detect them. Detecting the reconnection points could be challenging, particularly in the weak magnetic field case where the mixing layer is turbulent, resulting in numerous reconnection events \citep{Servidio2010}. Hence, an automated reconnection detection technique is necessary. While a reconnection detection algorithm already exists \citep{Servidio2010} (see $\S$\ref{sec:previous}), it is limited to 2D or quasi-2D (3D system with a strong guide field) systems. Hence, we develop a new reconnection detection algorithm that can be extended to 3D, see $\S$\ref{sec:new}. However, in this paper, we present the preliminary case of detecting reconnection points for a 2D system.

\vspace{-10pt}
\subsubsection{Previous reconnection detection techniques} \label{sec:previous}
\vspace{-10pt} 
In 2D, reconnection typically occurs at magnetic null points, which can exhibit either saddle (X-point) or center (O-point) topologies \citep{Lapenta_2021}. \citet{Servidio2010} proposed a detection algorithm based on this, originally for homogeneous 2D turbulence. Their method first identifies candidate points where the magnitude of the magnetic field strength is near zero. At these points, the eigenvalues of the Hessian matrix $\left(\partial^2 \star/\partial x_i \partial x_j \right)$ of magnetic vector potential $\mathcal{A}$ are calculated to gain insight into the magnetic field topology. The points with positive and negative eigenvalues have saddle topology, i.e., X-points. The above method was also used for the quasi-2D system \citep{Zhdankin_2013}, by detecting the reconnection points at each 2D plane along the guide field direction.

However, this approach is unsuitable for fully 3D configurations, where reconnection does not always occur at null points. Even in 2D, since eigenvalues were used to classify the X- and O-points, we must be as close to the real X or O-point as possible. Hence, this method demands an extremely high resolution or an accurate interpolation technique to map the data to finer scales to locate the X-points.

\vspace{-10pt}
\subsubsection{Present reconnection detection technique} \label{sec:new}
\begin{figure}
     \centering
     \begin{subfigure}[b]{\linewidth}
         \centering
         \includegraphics[width=0.67\linewidth]{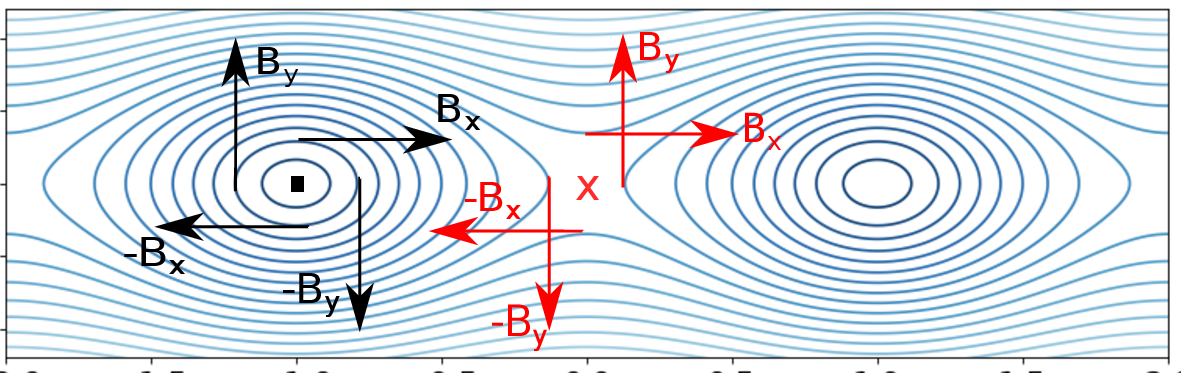}
     \end{subfigure}
     \hfill
     \begin{subfigure}[b]{\linewidth}
         \centering
         \includegraphics[width=0.67\linewidth]{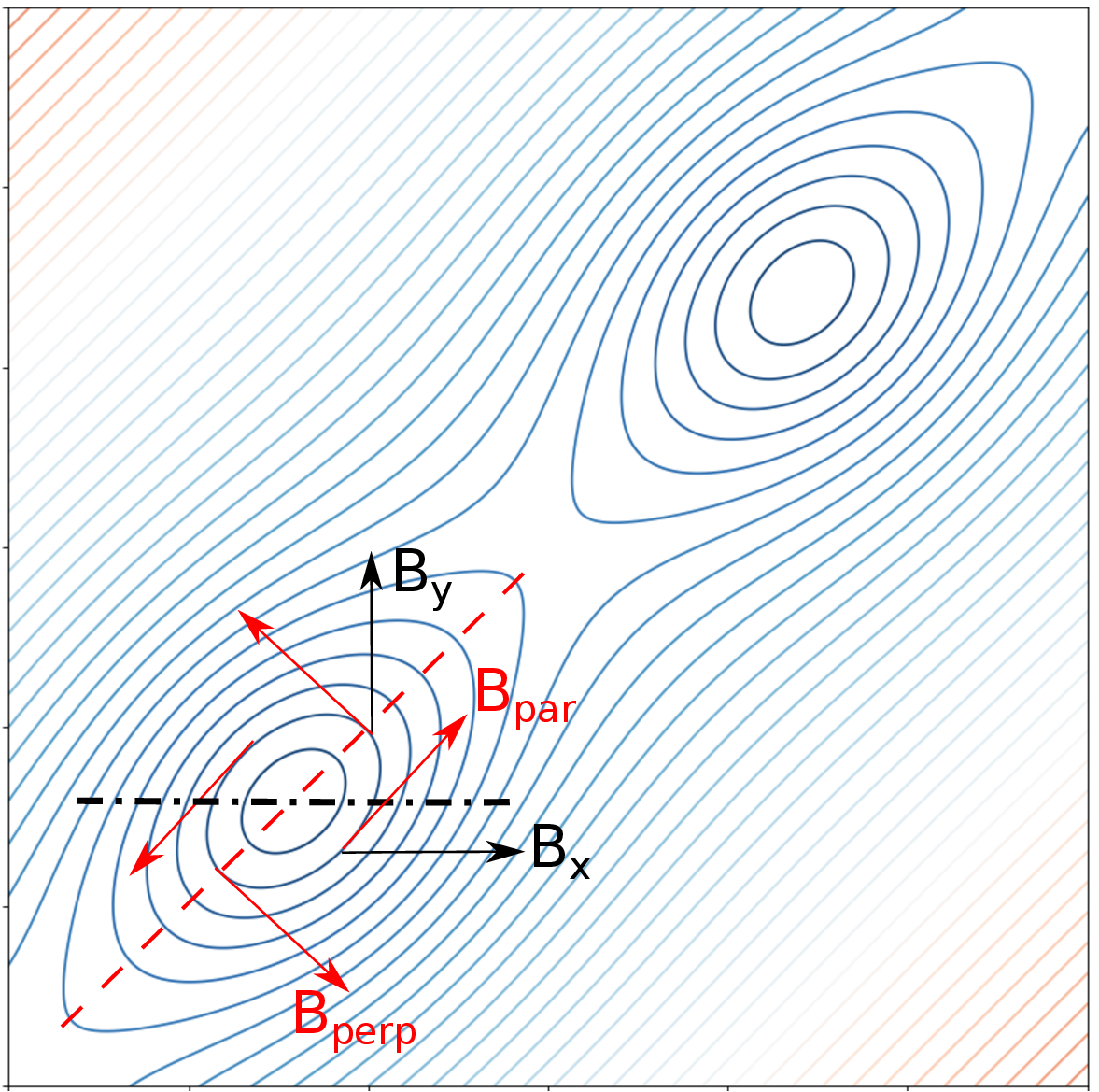}
     \end{subfigure}
     \caption{Magnetic field topology around X, O-points in a Fadeev equilibrium configuration aligned parallel (top), inclined (bottom) relative to the reference frame. The X and O-points are represented as $\times$} and \textcolor{black}{$\blacksquare$}.
     \label{x_o_points}
\end{figure}
\begin{figure*}
    \centering
    \includegraphics[width = \textwidth]{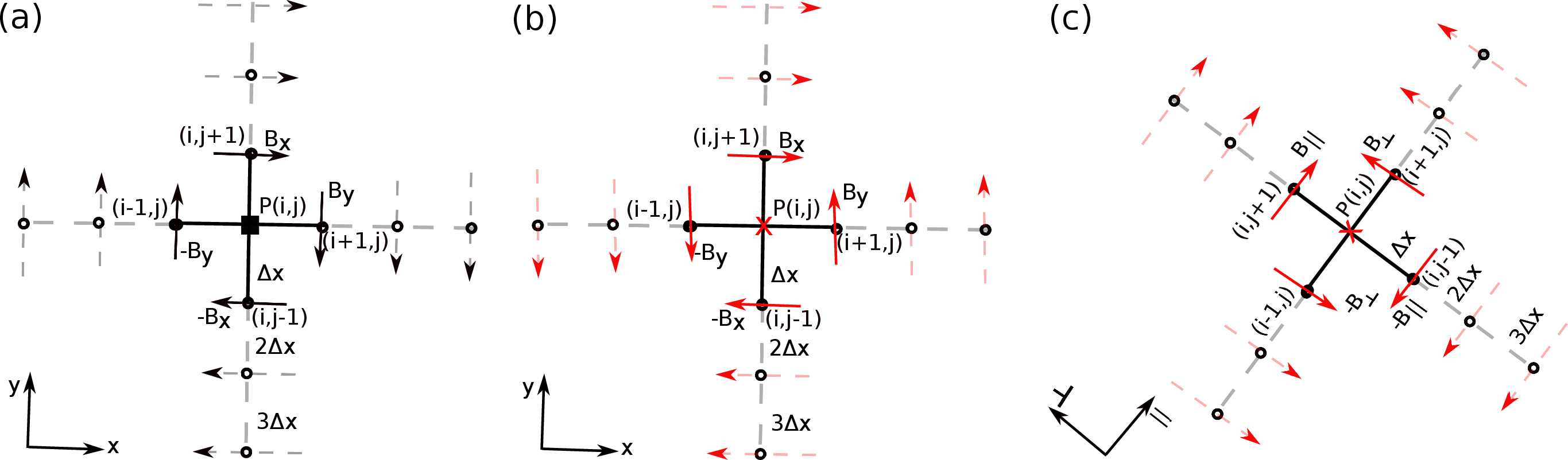}
    \caption{Schematic representation of five point stencil of different radii ($\Delta x, 2\Delta x, 3 \Delta x$) from the point under consideration $P(i,j)$. The schematics (a), (b) represent the magnetic field configuration for O-, X-points, respectively, when the frame of reference is parallel to the coordinate frame. (c) represent the magnetic field configuration for X-point when the frame of reference is inclined at an angle with respect to the coordinate frame.}
    \label{stencil}
\end{figure*}
\begin{figure*}
    \centering
    \includegraphics[width = \textwidth]{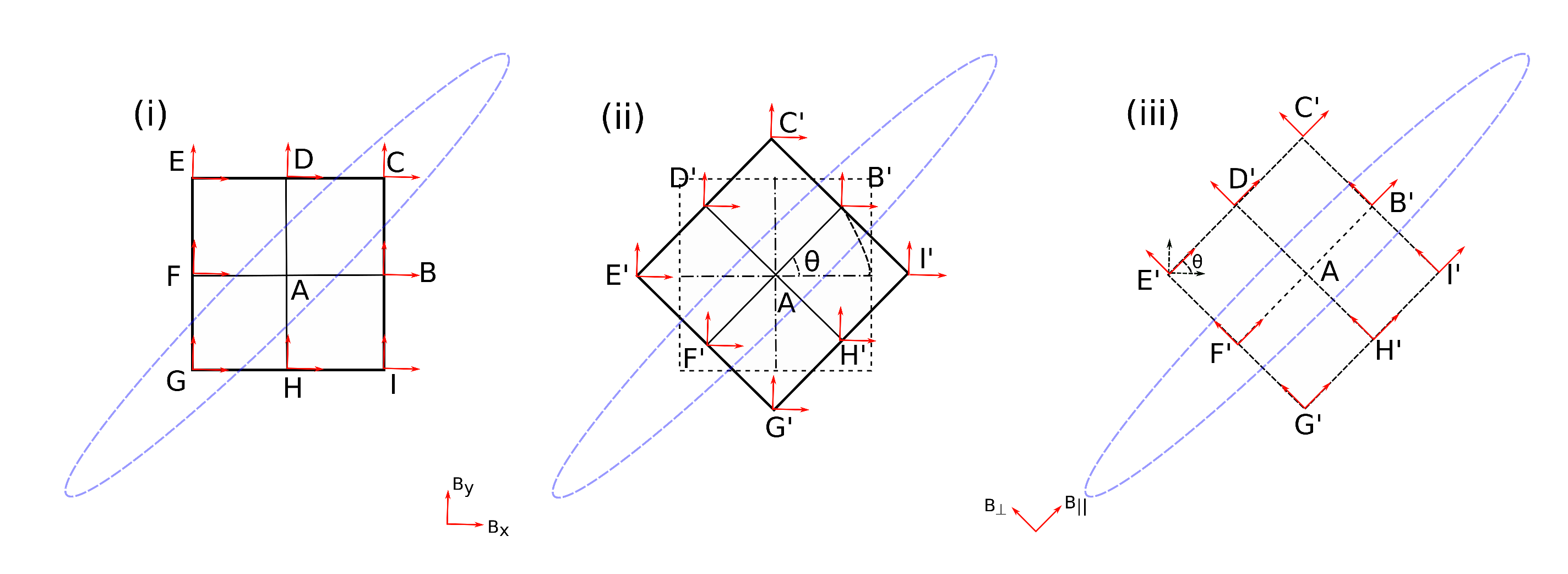}
    \caption{Schematic representation of the stages of calculating the components of the magnetic field parallel and perpendicular to the current sheet, along with an elongated plasmoid (blue ellipse) oriented at an angle $\theta$ to the coordinate axis.}
    \label{grid_rotation}
\end{figure*}

\vspace{-10pt} 
To overcome the limitations of 2D-based methods, we develop a generalized algorithm that detects reconnection points based on local magnetic field configurations. For simplicity, we will demonstrate the algorithm using the Fadeev equilibrium configuration. The Fadeev equilibrium \citep{Fadeev_1965, Yoon_2005} is a continuous arrangement of magnetic islands (plasmoids) with consecutive saddle and center topology leading to the X- and O-points (represented as $\times$ and \textcolor{black}{$\blacksquare$} respectively in figure \ref{x_o_points}). The Fadeev equilibrium is modelled as 
\begin{equation}
    \mathcal{A} = \frac{B_0}{\kappa } \log[\cosh(\kappa y) + \epsilon \cos(\kappa x)],
    \label{fadeev_equation}
\end{equation}
where $\mathcal{A}$ is the magnetic vector potential, $B_0$ is the initial magnetic field strength, and $\kappa$ represents the wave number. $\kappa$ determines the number of plasmoids in the configuration and the number of null points. For $\kappa {=} n \pi,$ we have $2n$ X- and O-points. $\epsilon$ controls the eccentricity of the plasmoid. The limits $\epsilon \rightarrow 0$ and $\epsilon \rightarrow 1$ represent the highly eccentric plasmoid and perfectly circular plasmoid, respectively. The limit $\epsilon = 0$ gives the Harris equilibrium, where the magnetic field lines are parallel and have no X- and O-points.
 
From figure \ref{x_o_points}(top), we see that the horizontal ($B_x$) and vertical ($B_y$) components of the magnetic field change sign across the reconnection point along the vertical ($y$) and horizontal ($x$) directions, respectively. To detect such points, we consider a 2D five-point stencil of radius $\Delta x$ around the point $P(i,j)$ (see the solid lines in figure \ref{stencil}), where $\Delta x$ is the grid resolution. We check if $B_y$ is changing along $x$ across the point $P$ using condition $B_y(i-1,j) {\times} B_y(i+1,j) {<} 0$. Similarly, we verify if $B_x$ is changing along $y$ using $B_x(i,j-1) {\times} B_x(i, j+1) {<} 0$. This is schematically shown in figure \ref{stencil} (a),(b). To avoid the null points due to noise, we discard points of order less than $-10^{-8}$. By filtering the points that satisfy both conditions, we obtain the center and saddle-type null points.

To extract the reconnection points alone, we use the cross gradients of magnetic field components $(\partial_x B_y, \partial_y B_x)$, where $\partial_{\star} \Phi = \partial \Phi/\partial \star$. The product of these cross gradients is positive for X-points and negative for O-type points. The points filtered above that satisfy the condition $\partial_x B_y {\times} \partial_y B_x > 0$ are taken as reconnection points. The magnetic field configuration at these reconnection points are further verified by repeating the above process with a five-point stencil of radius $2 \Delta x$ and $3 \Delta x$ (see the dashed lines in figure \ref{stencil}). The points that show saddle nature at all these radii are only considered as X-points.

The above algorithm is suitable when the reconnecting field lines are parallel to the coordinate axes, as shown in figure \ref{x_o_points}(top). However, due to the turbulent nature of the mixing layer, often the reconnecting field lines are oriented at an angle to the reference frame (see figure \ref{x_o_points}(bottom)). In such cases, the 2D five-point stencil must first be aligned with the field lines before the above detection algorithm is implemented. Towards this, we first detect the orientation of current at the point ($\mathbf{- -}$ in figure \ref{x_o_points}(bottom)) to the reference coordinate axis ($\textcolor{black}{\textbf{-.}}$ line in figure \ref{x_o_points}(bottom)). Since the current ($J$) is scalar quantity in 2D, we calculate the orientation of the current sheet using $\theta = \tan^{-1} \left( \partial_x J/\partial_z J \right)$. A $3 {\times} 3$ box is considered (figure \ref{grid_rotation}(i)) and the box is then oriented along the current sheet $\theta$ as shown in figure \ref{grid_rotation}(ii). The magnetic field components parallel and perpendicular to the current sheet (represented as $B_{\parallel}$ and $B_{\perp}$) are calculated (see figure \ref{grid_rotation}(iii)). We now check if the $B_{\parallel}, B_{\perp}$ are changing sign across the point of investigation using $B_{\perp}(i-1,j) {\times} B_{\perp}(i+1,j) < 0$, $B_{\parallel}(i,j-1) {\times} B_{\parallel}(i, j+1) < 0$ (figure \ref{stencil}(c)). We then filter the X-point based on the condition $\partial_{\parallel} B_{\perp} {\times} \partial_{\perp} B_{\parallel} > 0$. The issues of this method and their solutions were presented in $\S$\ref{Detection}. The reconnection detection algorithm was developed as a post-processing code. The current algorithm works only for the 2D case. 

\vspace{-10pt}
\subsection{Reconnection events in MRTI} \label{sec:number}
\begin{figure*}
     \centering
     \begin{subfigure}[b]{0.33\textwidth}
         \centering
         \includegraphics[width=\textwidth]{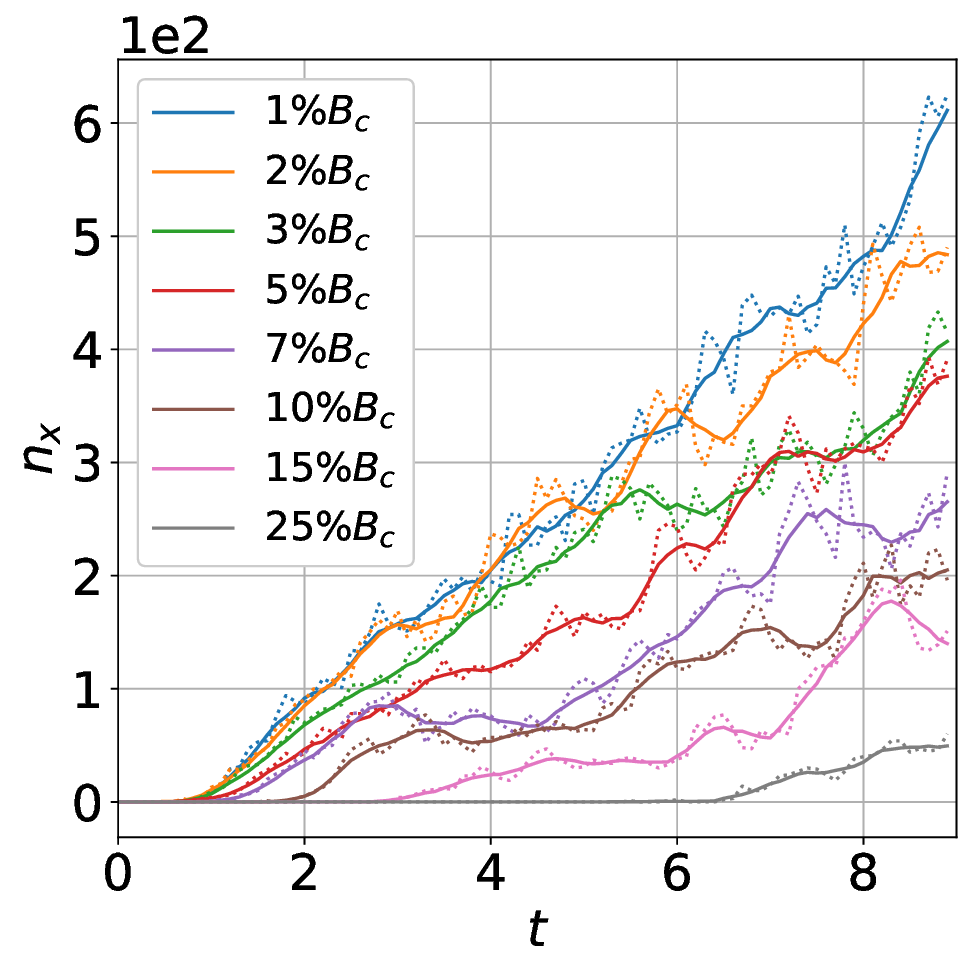}
     \end{subfigure}
     \hfill
     \begin{subfigure}[b]{0.31\textwidth}
         \centering
         \includegraphics[width=\textwidth]{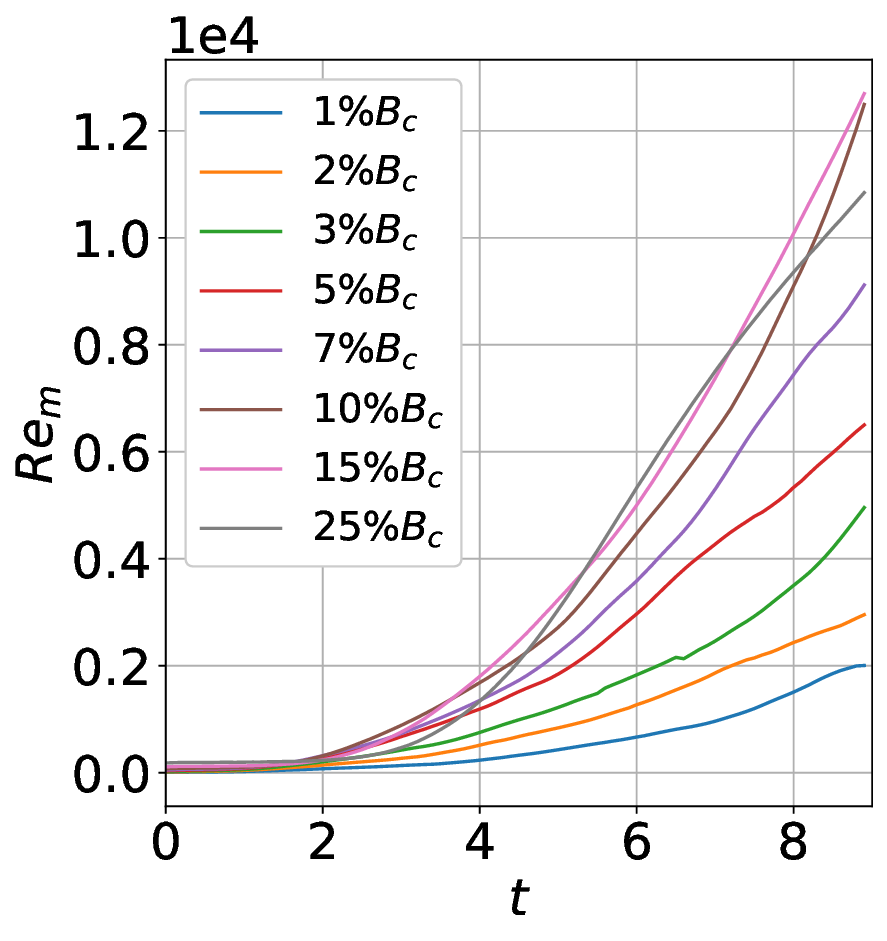}
     \end{subfigure}
     \hfill
     \begin{subfigure}[b]{0.33\textwidth}
         \centering
         \includegraphics[width=\textwidth]{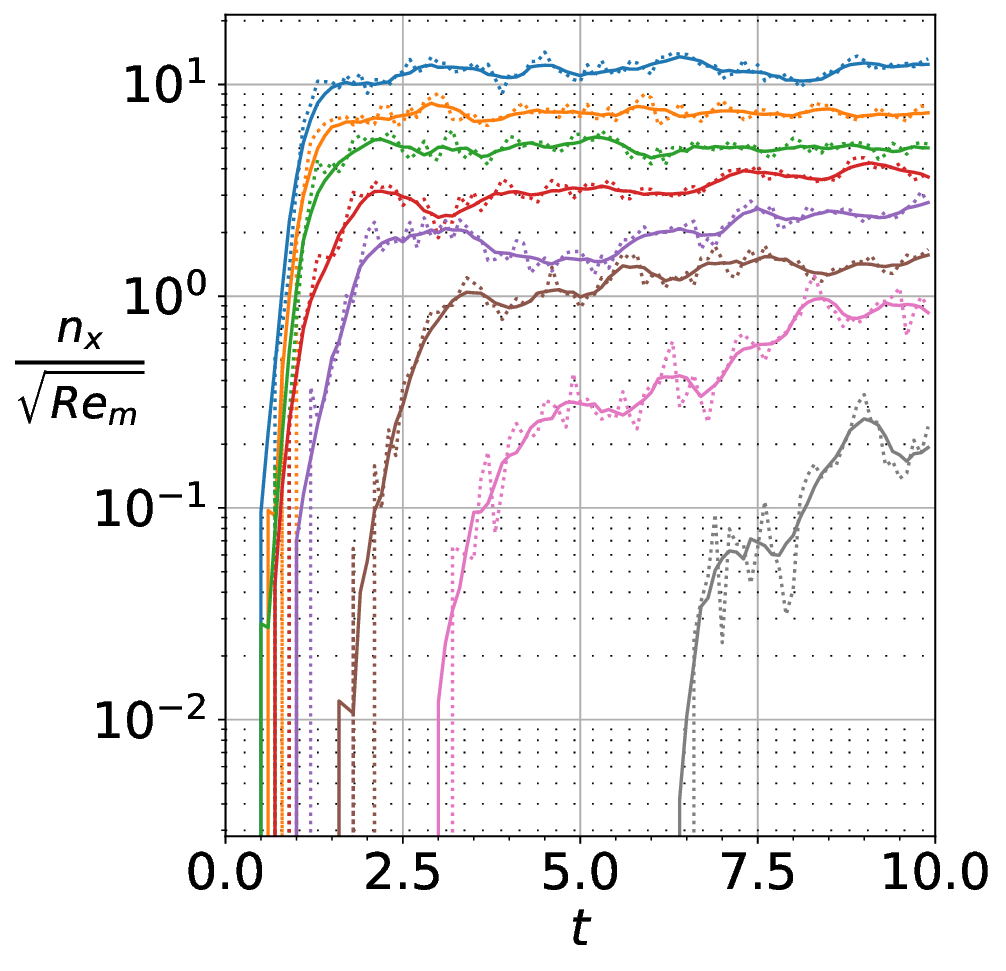}
     \end{subfigure}
     \caption{Figure showing the temporal variation of: \textit{(left)} number of reconnection events; \textit{(center)} magnetic Reynolds number ($Re_m$); \textit{(right)} number of reconnection events scaled with $\sqrt{Re_m}$. The quantity is plotted for different magnetic field strengths. In each case, the dotted and solid lines show the raw data and rolling average, respectively. The rolling average is taken with a window of $0.3 \tau$.}
     \label{nx_Re_t}
\end{figure*}

\vspace{-10pt} The developed algorithm is used to detect the reconnection points in the MRTI mixing layer for different magnetic field strength cases. From figure \ref{nx_Re_t}\textit{(left)}, we see that, for all magnetic field strengths, the number of reconnection points ($n_{\times}$) increases with time. The reason for the increase in $n_{\times}$ is attributed to two factors --- $i)$ height of the mixing layer and $ii)$ the turbulent motion of magnetic field lines. As the mixing layer widens and the turbulence increases, we see that the number of reconnection points increases. In the current work, we find that the number of reconnection points scales as the square root of the magnetic Reynolds number, as shown in figure \ref{nx_Re_t}(right). Potential reasons for this scaling law are discussed in $\S$\ref{nxexplanation}. From figure \ref{nx_Re_t}\textit{(right)} we see that for a given $Re_m$, $n_{\times}$ decreases with increasing magnetic field strength. For the current set of simulations, we find that 
\begin{equation}
    \frac{n_{\times}}{\sqrt{Re_m}} = \frac{0.15 B_c}{B_0}.
\end{equation}

\begin{figure*}
    \centering
    \includegraphics[width=\linewidth]{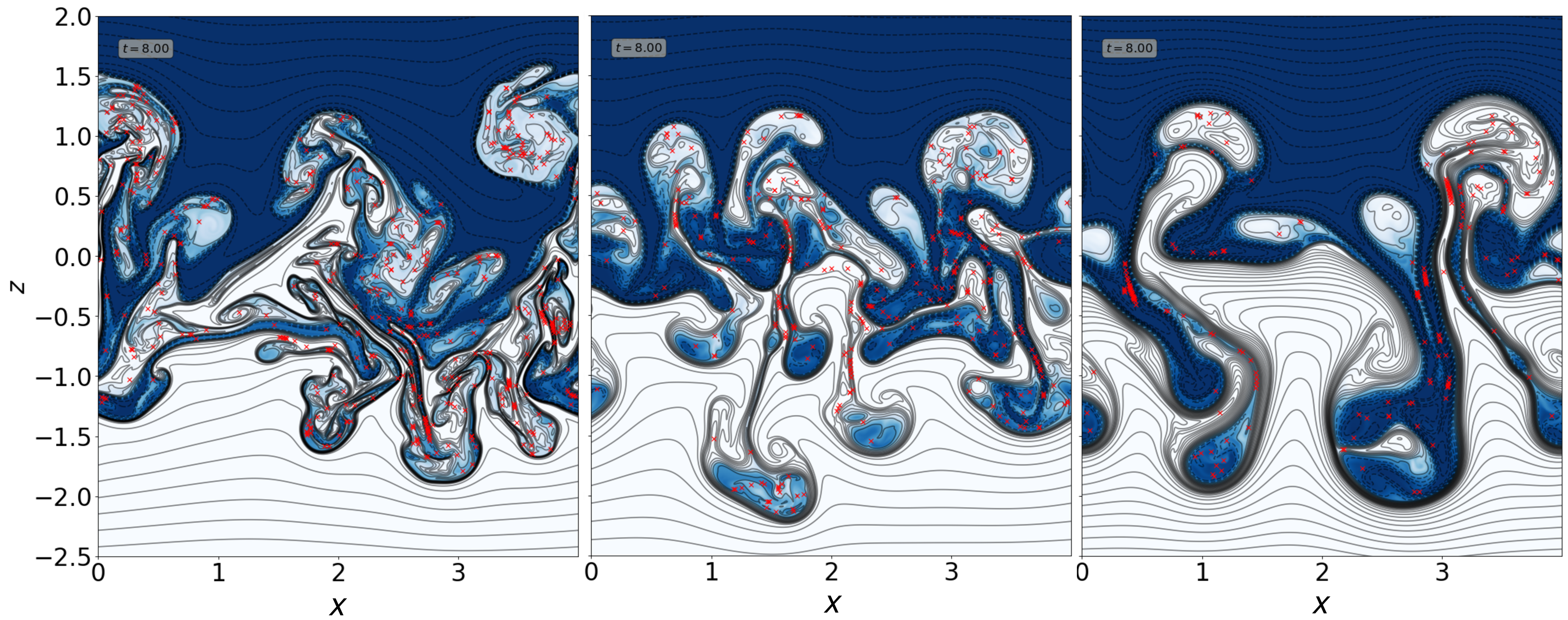}
    \caption{Figure showing the density contours of MRTI along with the magnetic field lines (black lines) and the reconnection points ($\times$) detected from the algorithm described in $\S$\ref{section:detection} at different magnetic field strengths: $B_0 {=} 1\%B_c$\textit{(left)}, $B_0 {=} 5\%B_c$\textit{(center)}, and $B_0 {=} 10\%B_c$\textit{(right)}. White and blue regions correspond to $\rho = 1, 3$ respectively.}
    \label{x_points}
\end{figure*}

To explain the above behaviour, we plot the instantaneous density contours of the mixing layers for different magnetic field strengths along with the detected reconnection points and magnetic field lines (see figure \ref{x_points}). The density contours show that the mixing layer is more turbulent at weaker field strengths. This is due to the selective suppression of small-scale perturbations ($k {\geq} k_c$) by the magnetic field in the linear regime \citep{Kalluri_2024}. The critical wave number $k_c$ is the smallest wave mode that can be suppressed by the parallel magnetic field in the linear regime \citep{Chandrasekhar1961}. From the linear growth rate equation, $k_c$ decreases with increasing magnetic field strength \citep{Chandrasekhar1961}. Thus, in the weak magnetic field case, a wide range of waves ($k {\in} [1, k_c]$) grow, and the mixing layer is turbulent \citep{Stone2007a, Carlyle2017, Kalluri_2024}. The turbulent nature of the mixing leads to frequent interaction of field lines, resulting in an increased number of reconnection points in the weak magnetic field case (see figure \ref{x_points}(left)). Conversely, $k_c$ is small in the strong magnetic field case, and only a limited small $k$ modes grow. Therefore, the mixing layer has laminar-like plumes (figure \ref{x_points}(right)). Due to less turbulence, the reconnection points are also low.  

\vspace{-10pt}
\subsection{Strength of the reconnection events} \label{sec:characteristics}
\begin{figure*}
    \centering
    \includegraphics[width = \textwidth, height = 0.33\textheight]{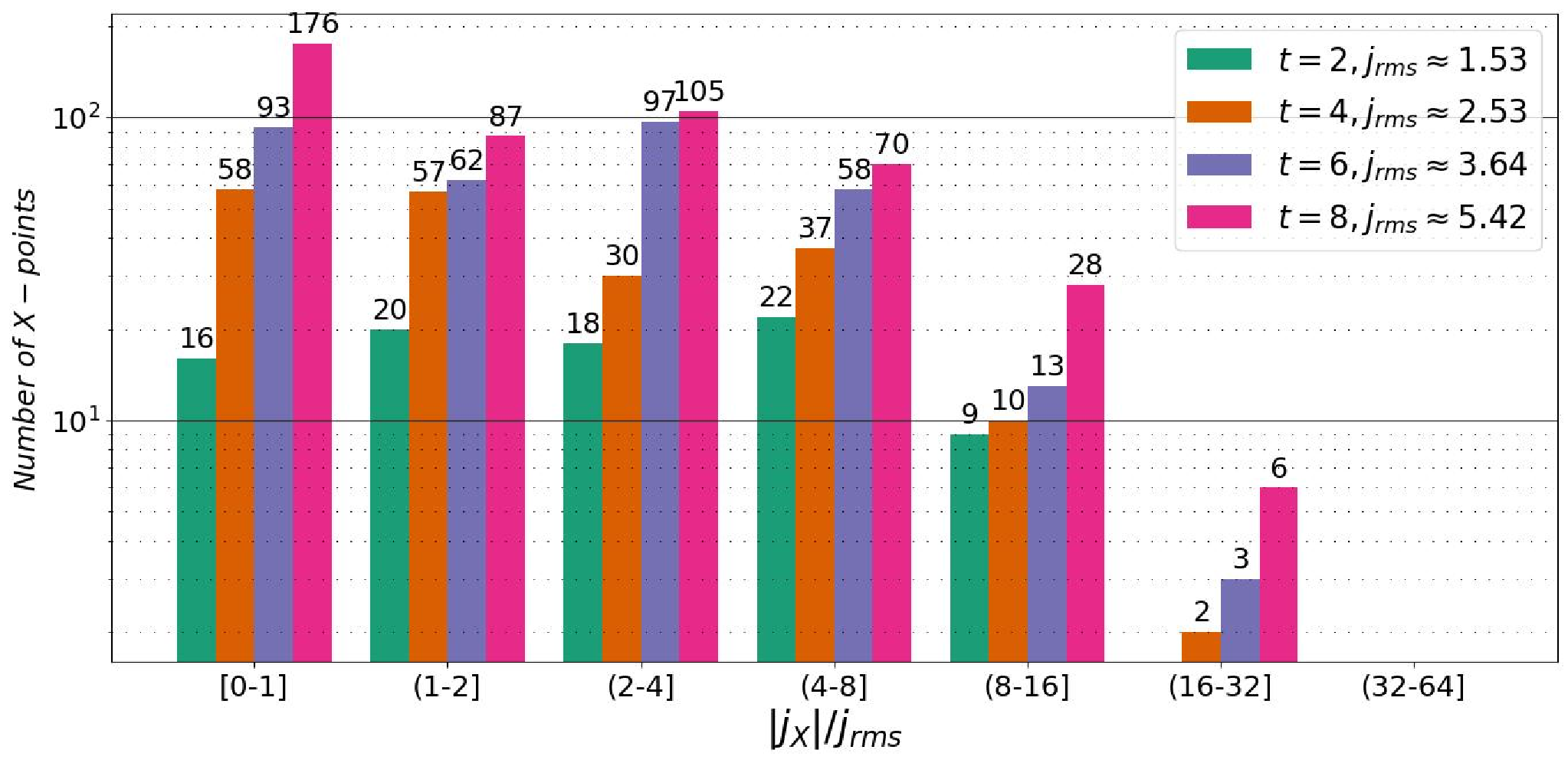} 
    \caption{Histogram plot showing the number of X-points and the corresponding strength of current sheets. The distribution is plotted for the case of $B_0 = 1\% B_c$.}
    \label{hist1}
\end{figure*}
\begin{figure*}
    \centering
    \includegraphics[width = \textwidth, height = 0.33\textheight]{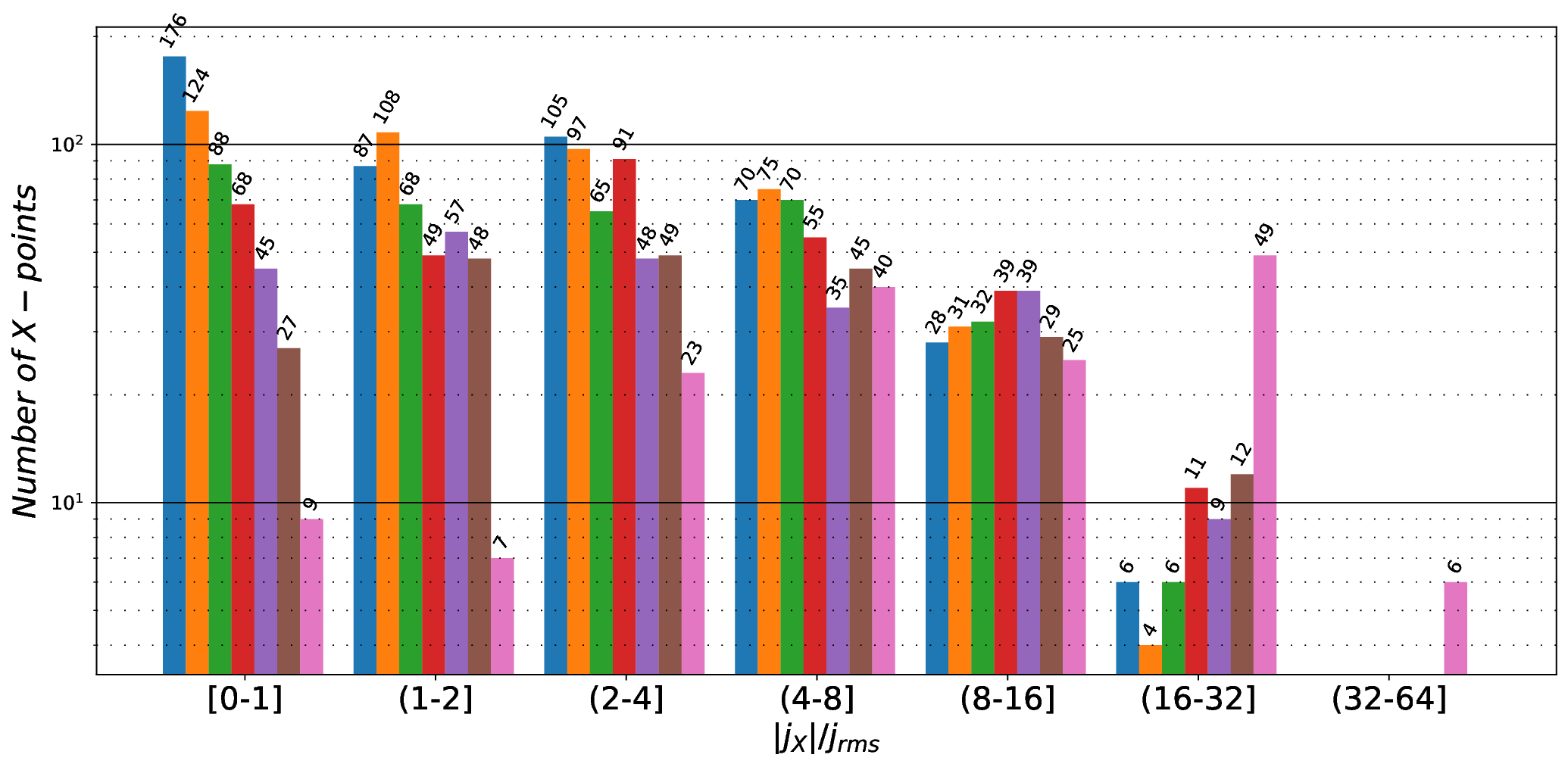}
    \caption{Distribution of current sheets leading to reconnection for different magnetic field strengths. The distribution is plotted for a given time instant $t = 9.8 \tau$. The colour of each magnetic field case is the same as figure \ref{nx_Re_t}.}
    \label{hist}
\end{figure*}

\vspace{-10pt} In the previous subsection, we quantified the reconnection events. Here, we will delve into their characteristics, particularly the strength of the reconnecting current sheets. This provides a quantitative insight into the influence of reconnection on energy dynamics, like energy conversion to kinetic energy and energy dissipation. Further, the variation of these characteristics with time and imposed magnetic field strength is discussed. 

The first step is to segregate the current sheets based on their strength. However, due to the increase in turbulence and magnetic field amplification, the nature of the mixing layer and the characteristics of current sheets change with time. Similarly, due to suppression of perturbations, the nature of the mixing layer (and in turn the characteristics of the current sheets) changes with magnetic field strength. Therefore, a direct comparison of current sheets across time and magnetic field strengths is inappropriate. To make the reconnecting current sheets comparable across different time instants and magnetic field strengths, we need to normalize the current sheet by the instantaneous root mean square current $\left( j_{rms} = \int_{V} |j| \mathrm{d}V \right)$, where $V$ is the volume of mixing layer. As the instability evolves, the height of the mixing layer and the strength of current sheets increase. Capturing these, $j_{rms}$ also increases (see the legend of figure \ref{hist1}), making it an appropriate normalization parameter. 

The normalized reconnecting current sheets are segregated into seven bins: current sheets that are weaker than $j_{rms}$ and then on in powers of 2 ($2^n$ to $2^{n+1}$, $n \in [0, 5]$). Figure \ref{hist1} shows the histogram of reconnecting current sheets in each range at various time instants. The histogram is plotted for the field strength $B_0 = 1\% B_c$. The proportion of strong current sheets (hereon, defined as the current sheets with strength greater than $8 j_{rms}$) is observed to be less than $\approx 7\%$ of the total reconnecting current sheets at any well-developed non-linear state ($t > 4$). At $t = 4, 6, 8$, the percentage of strong reconnecting current sheets is approximately $6\%,$ $5\%,$ and $7\%$, respectively. The choice of the threshold $> 8 j_{rms}$ for considering as strong current sheet comes from previous studies where they considered $5 j_{rms}$ \citep{Zhdankin_2013}, $10  j_{rms}$ \citep{Kadowaki_2018}. However, the choice is arbitrary, and quantitative results vary accordingly, while the broad conclusions remain the same. 

The characteristics of the reconnecting current sheets also vary with the magnetic field strength. Figure \ref{hist} shows the distribution of current sheets for MRTI of different magnetic field strengths at a well-evolved non-linear state, $t = 8$, see figure \ref{x_points}. The proportion of strong reconnecting sheets is increasing with field strength. The percentage of strong reconnecting sheets is approximately $7\%,$ $8\%,$ $11.5\%,$ $16\%,$ $20.5\%,$ $19.5\%,$ and $50\%$ for $B_0 = 1,$ $2,$ $3,$ $5,$ $7,$ $10,$ and $15$ $(\% B_c)$ respectively. This trend suggests that while reconnection is more frequent in weak-field, turbulent flows, it is more intense in strong-field, laminarized regimes, reflecting the concentration of magnetic energy into fewer, more coherent structures.

\vspace{-10pt} \subsection{Section summary}

\vspace{-10pt} To summarize $\S$\ref{Characteristics}, an algorithm that detects the reconnection points based on magnetic field configuration was developed. Using the algorithm, we quantified the number of reconnection points in the mixing layer as the instability evolves for different magnetic field strengths. For all field strengths, the number of reconnection points ($n_{\times}$) increases with time in a self-similar fashion, following $n_{\times} {\propto} t^{3/2}$. The proportionality constant decreases with increasing magnetic field strength. The large number of reconnection points in the weak field case could be attributed to the higher turbulence in the system. Analysis of current sheet strength revealed that most reconnection events are relatively weak ($< 8 j_{rms}$), but the fraction of strong events increases with magnetic field strength. The study highlights a key contrast between the reconnection in weak and strong magnetic field cases: while the former is spatially distributed and frequent, the latter is more localized but energetically significant.

\vspace{-10pt}
\section{Energy dynamics due to reconnection: Quantitative} \label{sec:energydynamicsquant}

\begin{figure*}
    \centering
    \includegraphics[width = \linewidth]{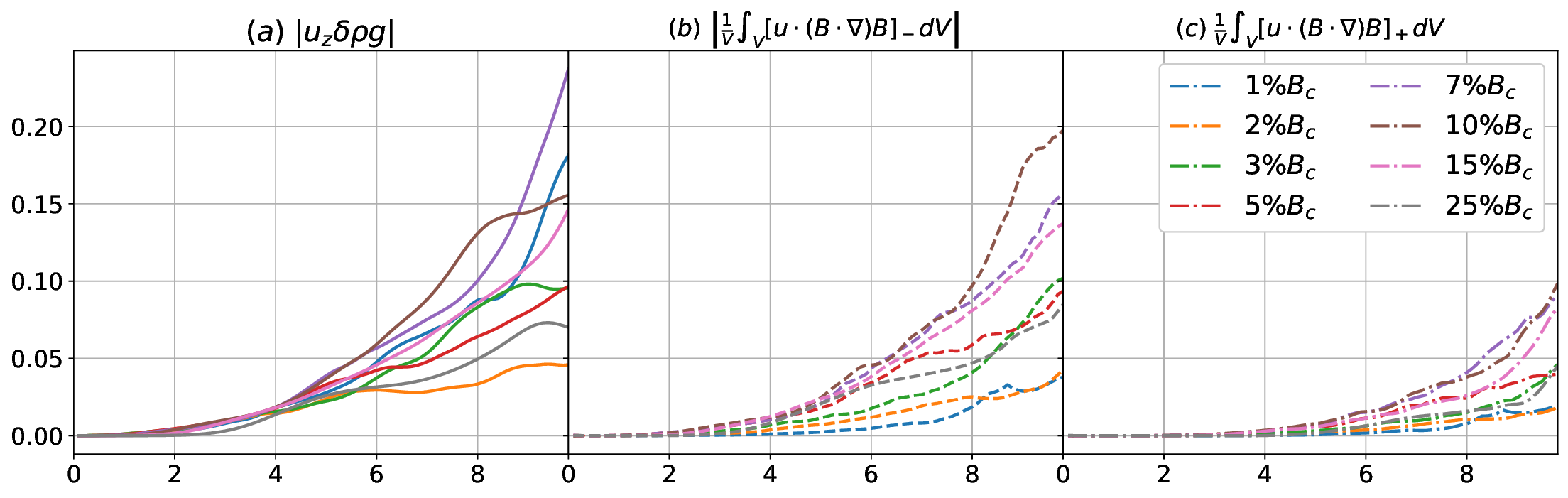}
    \caption{Temporal variation of: (a) the energy extracted from the GPE, given by $u_z \delta \rho g$; (b) energy transfer from fluid to magnetic field $[\mathbf{u} \cdot (\mathbf{B} \cdot \nabla) \mathbf{B}]_-$; and (c) energy transfer from magnetic field to fluid $[\mathbf{u} \cdot (\mathbf{B} \cdot \nabla) \mathbf{B}]_+$.}
    \label{figubdbquant}
\end{figure*}

In $\S$\ref{section:pinching} we illustrated that magnetic reconnection controls the TME by relieving magnetic tension. In $\S$\ref{sec:energy_dynamics}, we illustrated that the reconnection converts magnetic energy partly to kinetic energy, dissipating the rest. However, a quantitative analysis was deferred due to the lack of precise reconnection point data. Having developed a reconnection detection algorithm (see $\S$\ref{section:detection}), we are equipped to quantify the role of reconnection in the system’s energy dynamics, which is the focus of this section.

Before we delve into the energy dynamics due to reconnection, let us understand the energy flow in the MRTI system. Energy enters the mixing layer as fluid extracts gravitational potential energy (GPE), which increases the total kinetic energy (TKE). This energy extraction is estimated by the quantity $u_z \delta \rho g$. A portion of this kinetic energy is transferred into the magnetic field (see blue regions in figure \ref{ubdb_contour}), increasing the TME. Some of this magnetic energy is then returned to the fluid (see red regions in figure \ref{ubdb_contour}). This exchange between magnetic and kinetic energy is described by the term $\mathbf{u} \cdot (\mathbf{B} \cdot \nabla) \mathbf{B}$, as discussed in $\S$\ref{energyinterplay}. To isolate directional energy transfer, we define $[\mathbf{u} \cdot (\mathbf{B} \cdot \nabla) \mathbf{B}]_-$ for flow-to-field energy transfer, and $[\mathbf{u} \cdot (\mathbf{B} \cdot \nabla) \mathbf{B}]_+$ for field-to-flow transfer. We analyze these terms by computing volume-averaged values of $u_z \delta \rho g$, $[\mathbf{u} \cdot (\mathbf{B} \cdot \nabla) \mathbf{B}]_-$, and $[\mathbf{u} \cdot (\mathbf{B} \cdot \nabla) \mathbf{B}]_+$ over time, as shown in figure \ref{figubdbquant}. The results show that most GPE extraction (panel a) is transferred into the magnetic field (panel b), and about half of this energy is subsequently returned to the fluid (panel c). Consequently, the TKE is modulated by $[\mathbf{u} \cdot (\mathbf{B} \cdot \nabla) \mathbf{B}]_+$, while the net TME is governed by the difference $\left( [\mathbf{u} \cdot (\mathbf{B} \cdot \nabla) \mathbf{B}]_- - [\mathbf{u} \cdot (\mathbf{B} \cdot \nabla) \mathbf{B}]_+ \right)$.

\vspace{-10pt}
\subsection{Role of reconnection in magnetic energy dissipation} \label{tmedissquant}

Reconnection contributes to the energy budget not only by transferring magnetic energy to the fluid but also by dissipating magnetic energy. This dual role was illustrated qualitatively in $\S$\ref{sec:energy_dynamics}. Now, having identified reconnection sites, we quantitatively assess their contribution. To estimate dissipation, we focus on the reconnecting current sheets. Below is a step by step procedure of how we estimate energy dissipation due to reconnection: 
\begin{enumerate}
    \item Define a region of size $60\Delta x \times 60\Delta z$ centered about reconnection point \footnote{Larger regions yielded similar results.}. 
    \item Rotate the region about the reconnection point (as shown in figure \ref{grid_rotation}) such that the orientation is aligned with current sheet. Orientation of current sheet ($\theta$) is computed as described in $\S$\ref{section:detection}.
    \item Compute the root-mean-square current $\Hat{j}_{rms}$ in the box. Note that the $\Hat{j}_{rms}$ specified here is different from $j_{rms}$, as the former represents the RMS current calculated from the local rotated box of $60 \Delta x \times 60 \Delta z$, while the latter is calculated from the whole domain.
    \item Starting from the reconnection point, we trace along and across the sheet until the current drops to $2 \Hat{j}_{rms}$, defining the sheet’s length and width \footnote{The choice of $2 \Hat{j}_{rms}$ comes from trial and error of various sampled reconnection points. We found that the threshold of $\Hat{j}_{rms}$ includes other regions of the patch beyond the reconnecting current sheet, and threshold larger than $2 \Hat{j}_{rms}$ are increasingly more stringent, neglecting the outer regions of reconnecting current sheets}. Because $\Hat{j}_{rms}$ is locally defined, this approach is robust for weak and strong reconnection alike.
\end{enumerate}
The above process is repeated for all reconnecting current sheets. The total energy dissipation is the volumetric average of energy dissipation from all reconnection events.

\begin{figure}
    \centering
    \includegraphics[width = 0.75\linewidth]{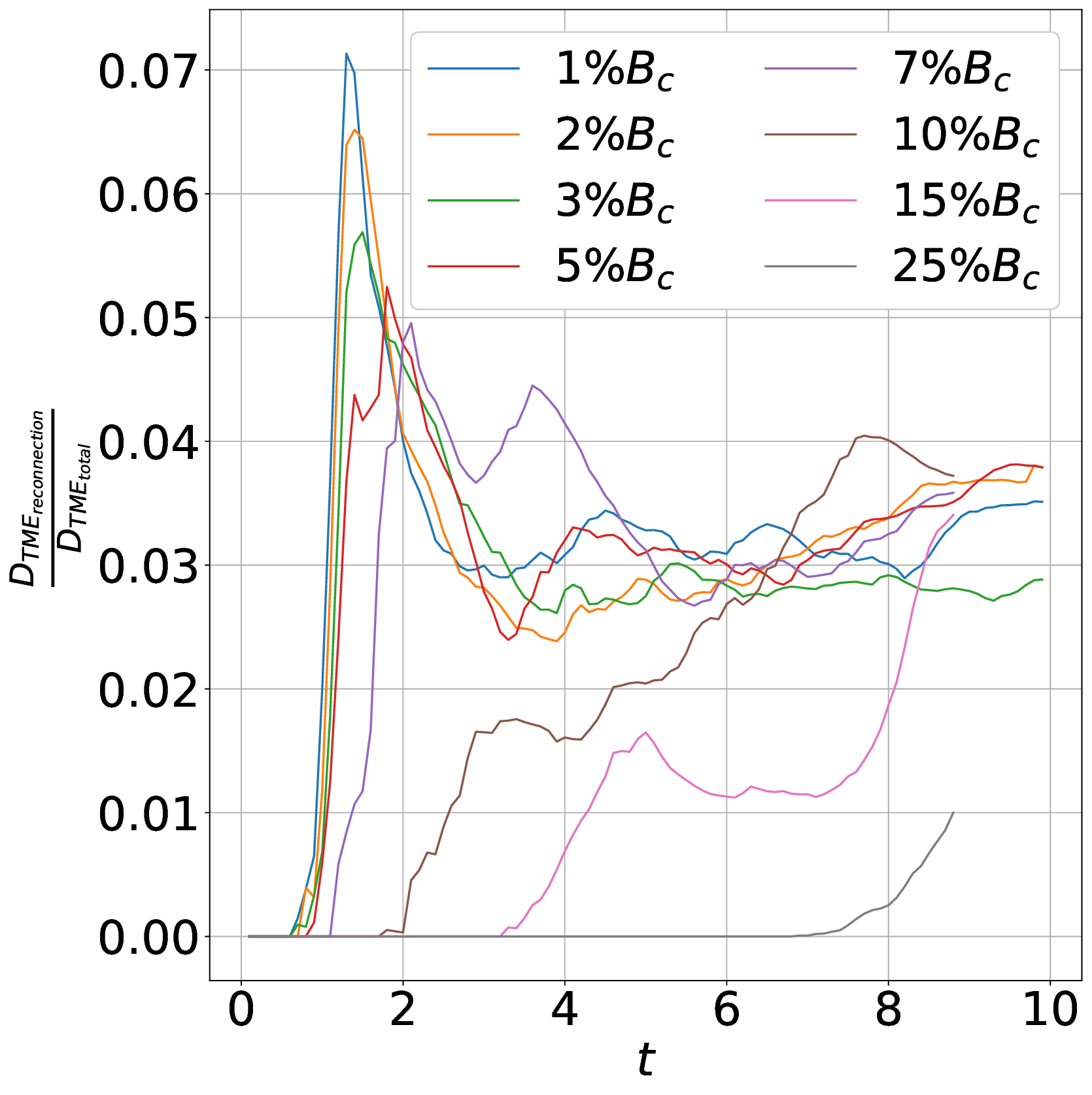}
    \caption{Temporal variation of the ratio of TME dissipation due to reconnection to the total TME dissipation for different magnetic field strengths}
    \label{mag_diss}
\end{figure}

The TME dissipation is computed as $\eta (\partial_j b_i)^2$ for each reconnecting current sheet based on the current sheet's dimensions. Integrating this dissipation in space and time over all reconnection events yields the total magnetic energy loss attributable to reconnection. We find that reconnection accounts for approximately $3\%$ of the total TME dissipation (see figure \ref{mag_diss}). Thus, while reconnection clearly contributes to dissipation, its overall role is relatively minor. This conclusion holds when current sheet dimensions are alternatively estimated using the Full Width at Half Maximum (FWHM) method, confirming the robustness of the result.

\vspace{-10pt}
\subsection{Role of reconnection in the energy transfer from magnetic field to fluid} \label{ubdbquant}

\begin{figure}
    \centering
    \includegraphics[width = 0.75\linewidth]{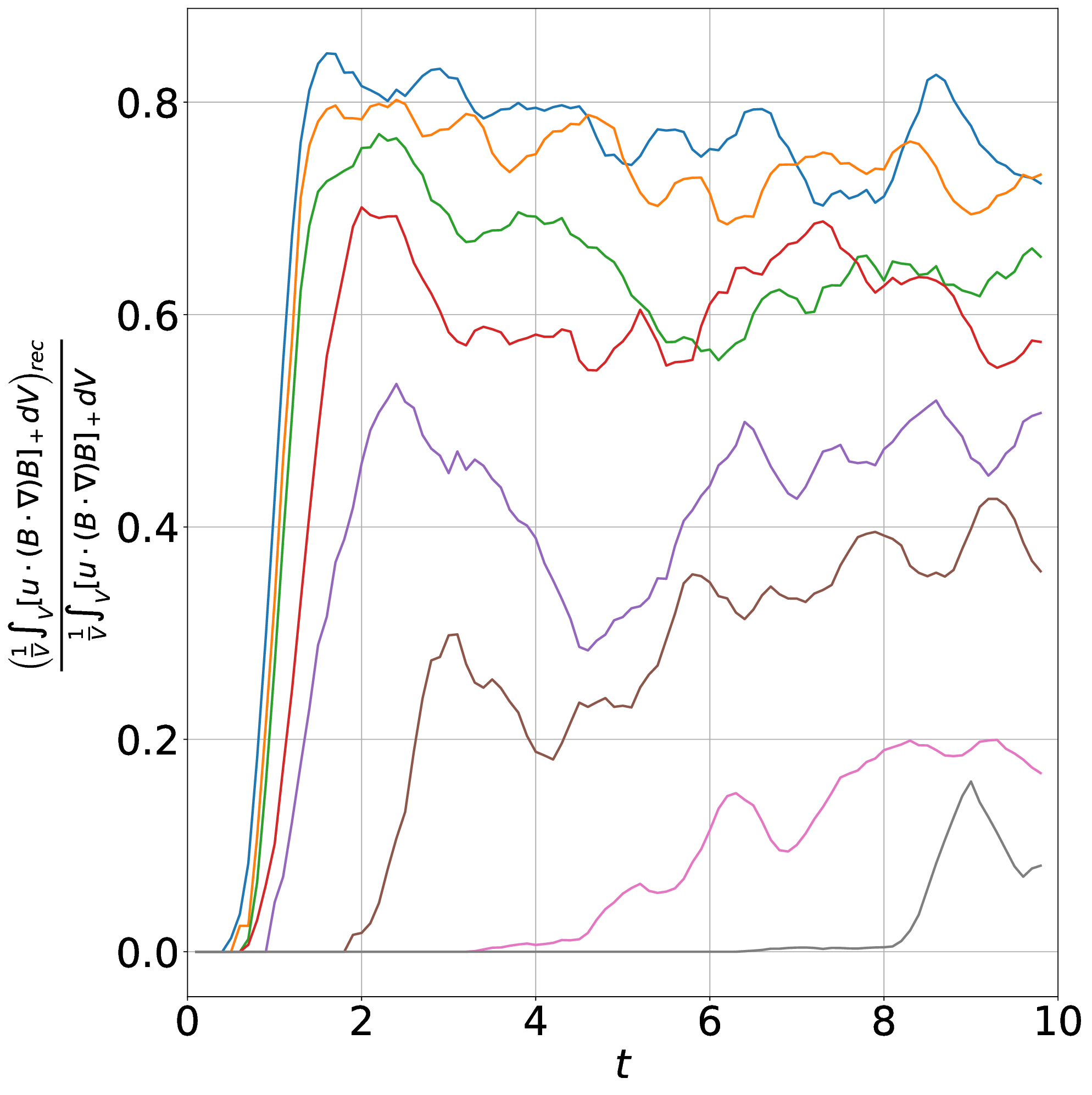}
    \caption{Temporal variation of the ratio of $[\mathbf{u} \cdot (\mathbf{B} \cdot \nabla) \mathbf{B}]_+$ due to reconnection to the total $[\mathbf{u} \cdot (\mathbf{B} \cdot \nabla) \mathbf{B}]_+$ for different magnetic field strengths. The data is smoothed by performing a rolling average with a window of $0.3$ time units. The legend is same as figure \ref{mag_diss}.}
    \label{ubdb_rec}
\end{figure}

Estimating the energy transferred from the magnetic field to the fluid is less straightforward, as it depends on the dynamics of reconnection outflows, which spread spatially, and the extent of spatial influence depends on the strength of the current sheet and local turbulence. To obtain a practical estimate, we define a box of size $100\Delta x \times 100\Delta z$ centered at each reconnection point, shown by dashed lines in figure \ref{boxubdb} (box size effects are discussed in $\S$\ref{sec:boxsize}). Within each box, we compute $[\mathbf{u} \cdot (\mathbf{B} \cdot \nabla) \mathbf{B}]_+$. Repeating this for all reconnection sites and averaging over volume gives the total contribution of reconnection to field-to-flow energy transfer. This is shown in figure \ref{ubdb_rec}. In contrast to energy dissipation, reconnection plays a significant role in energy transfer. At low magnetic field strengths, as much as $80\%$ of the energy transfer from the magnetic field to the fluid occurs near reconnection points, indicating that reconnection plays a dominant role in this regime. This correlates with the higher number of reconnection events observed in weak-field cases. As the magnetic field strength increases, the fraction of energy transfer attributable to reconnection decreases.

\subsubsection{Reconnection box size effect on the dynamics} \label{sec:boxsize}
\begin{figure}
    \centering
    \includegraphics[width= 0.9\linewidth]{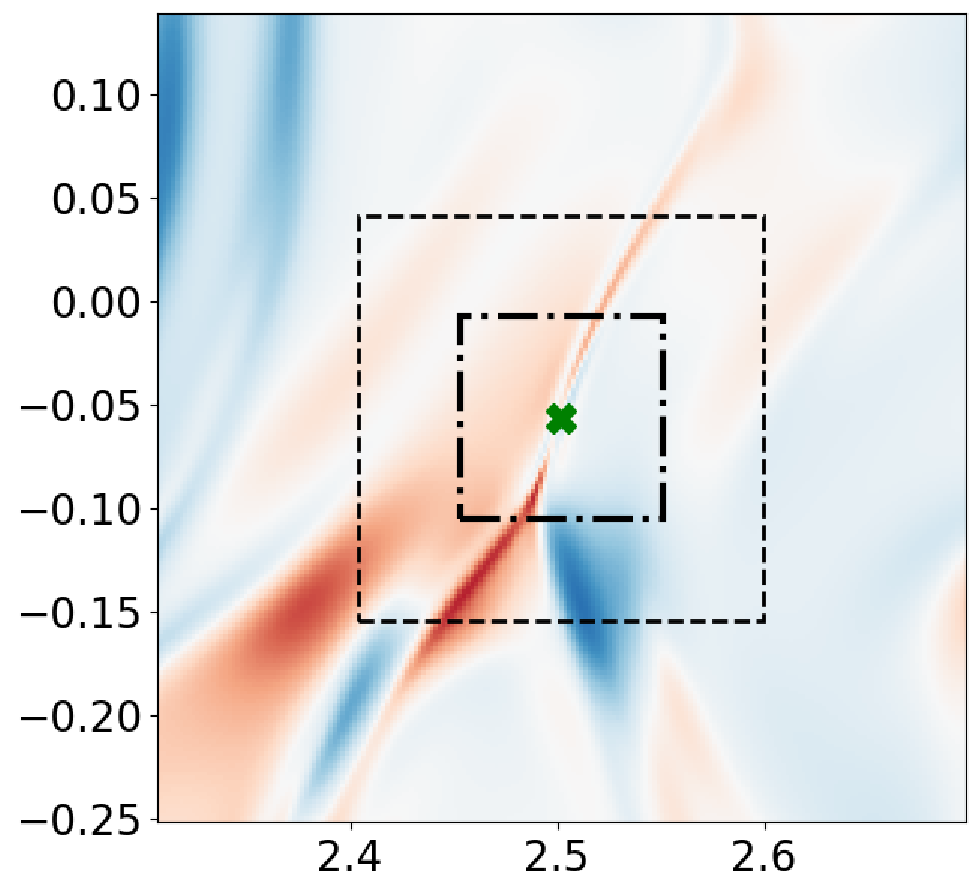}
    \caption{Contour of $\mathbf{u} \cdot (\mathbf{B} \cdot \nabla) \mathbf{B}$ in the box of $200 \Delta x \times 200 \Delta z$, $100 \Delta x \times 100 \Delta z$ (dashed), and $50 \Delta x \times 50 \Delta z$ (dashed-dotted) centered around a chosen reconnection point. The point is same as figure \ref{ubdb_contour}.}
    \label{boxubdb}
\end{figure}

To quantitatively estimate reconnection's influence on energy dynamics, we considered a box of $100 \Delta x \times 100 \Delta z$ in $\S$\ref{ubdbquant}. This is to ensure that the region of reconnection influence is well covered. Through figure \ref{boxubdb} we evidence that the considered box size encompasses the reconnection influenced region. A larger box will include the dynamics beyond the reconnection, and a smaller box does not capture the dynamics from reconnection fully. From figure \ref{boxubdb}, the outflow jets are hardly captured from $50 \Delta x \times 50 \Delta z$ box. Note that the same reconnection point chosen in figure \ref{ubdb_contour}(right) is considered here.

\begin{figure}
    \centering
    \includegraphics[width= 0.9\linewidth]{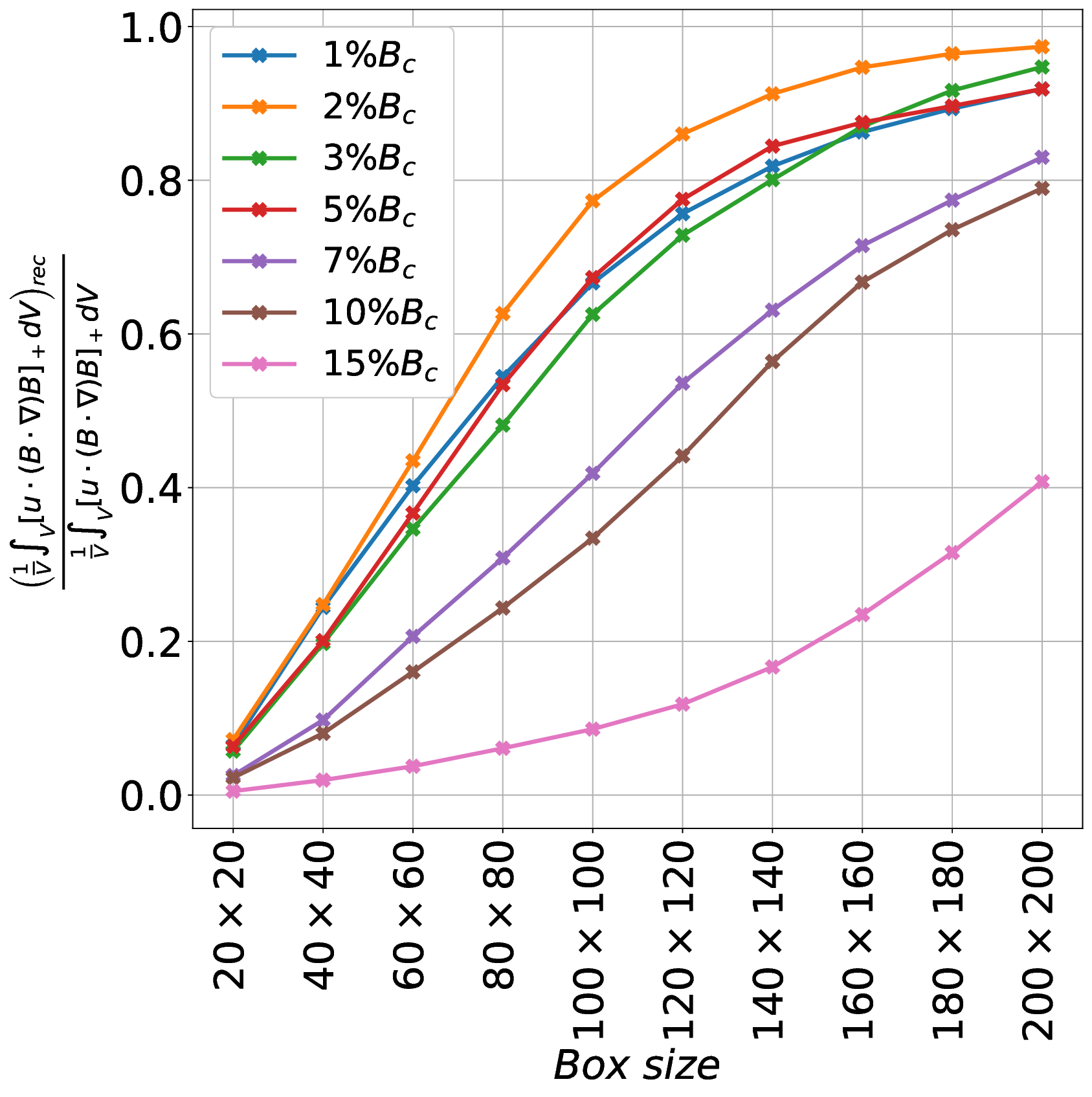}
    \caption{Contribution of magnetic reconnection to the total magnetic to kinetic energy conversion is estimated for different box sizes. The result is shown for different magnetic field strengths. The data for all cases is taken at the same time instant, $t=7$.}
    \label{boxsize}
\end{figure}

Evidently, the choice of box size does influence the magnitude of $\left( \frac{1}{V} \int_V [\mathbf{u} \cdot (\mathbf{B} \cdot \nabla) \mathbf{B}]_+ dV \right)_{rec}$. But, we acknowledge that considering a single box size may not be applicable for all reconnection points, given the wide range of characteristics of reconnecting current sheets (cf. figures \ref{hist}, \ref{hist1}). But, choosing a different box size for each reconnection depending on current sheet strength is tedious. Further, a appropriate relation between the current sheet strength and box size is unavailable. To quantitatively estimate the influence of box, we calculate $\left( \frac{1}{V} \int_V [\mathbf{u} \cdot (\mathbf{B} \cdot \nabla) \mathbf{B}]_+ dV \right)_{rec}$ at a particular time instant (t = 7) for different box sizes (X-axis) and magnetic field strengths. Normalizing with total $\frac{1}{V} \int_V [\mathbf{u} \cdot (\mathbf{B} \cdot \nabla) \mathbf{B}]_+ dV$ at the considered time instant (t = 7) and magnetic field strengths, we determine the contribution of reconnection to total magnetic to kinetic energy conversion. From figure \ref{boxsize}, we see that the box size has a linear relationship with $\frac{\left(\frac{1}{V} \int_V [\mathbf{u} \cdot (\mathbf{B} \cdot \nabla) \mathbf{B}]_+ dV\right)_{rec}}{\frac{1}{V} \int_V [\mathbf{u} \cdot (\mathbf{B} \cdot \nabla) \mathbf{B}]_+ dV}$.

\vspace{10pt}
Note that we use different box sizes to capture energy dissipation ($60 \Delta x \times 60 \Delta z$) and energy outflow ($100 \Delta x \times 100 \Delta z$), since the energy dissipation is more dominant at the reconnection point, as shown in figure \ref{mag_diss_contour}(right), while outflow jet dynamics are more dominant away from reconnection point, as shown in figure \ref{ubdb_contour}(right).

\vspace{-10pt}
\subsection{Reconnection rate}

\vspace{-10pt} Another characteristic of magnetic reconnection is the rate at which the field lines move through the reconnection point, i.e., how fast the reconnection occurs, called the reconnection rate. For a steady-state 2D reconnecting system, the reconnection rate is calculated based on the electric field ($E = u {\times} B + \eta j$) at the reconnection point. Since the reconnection point in 2D is a magnetic null point ($|B| \rightarrow 0$), the electric field reduces to $\approx \eta j$. 

In our system, reconnection occurs at many locations simultaneously, each with different current sheet properties. Consequently, reconnection proceeds at a range of rates—i.e., the system exhibits a distribution of reconnection rates rather than a single characteristic value. These distributions resemble those shown in figures \ref{hist1} and \ref{hist}, depending on whether we vary time at fixed magnetic field strength or vice versa, respectively.

Nevertheless, to understand how reconnection rate scales with global parameters such as the magnetic Reynolds number, it is useful to define an aggregate quantity. We compute the total reconnection rate $RR$ at each time instant by summing the reconnection contributions from all $n_{\times}$ detected reconnection events:

\begin{equation}
    RR = \sum_{i = 1}^{n_{\times}} \eta |j_i|.
    \label{RR}
\end{equation}

\begin{figure}
    \centering
    \includegraphics[width= 0.75\linewidth]{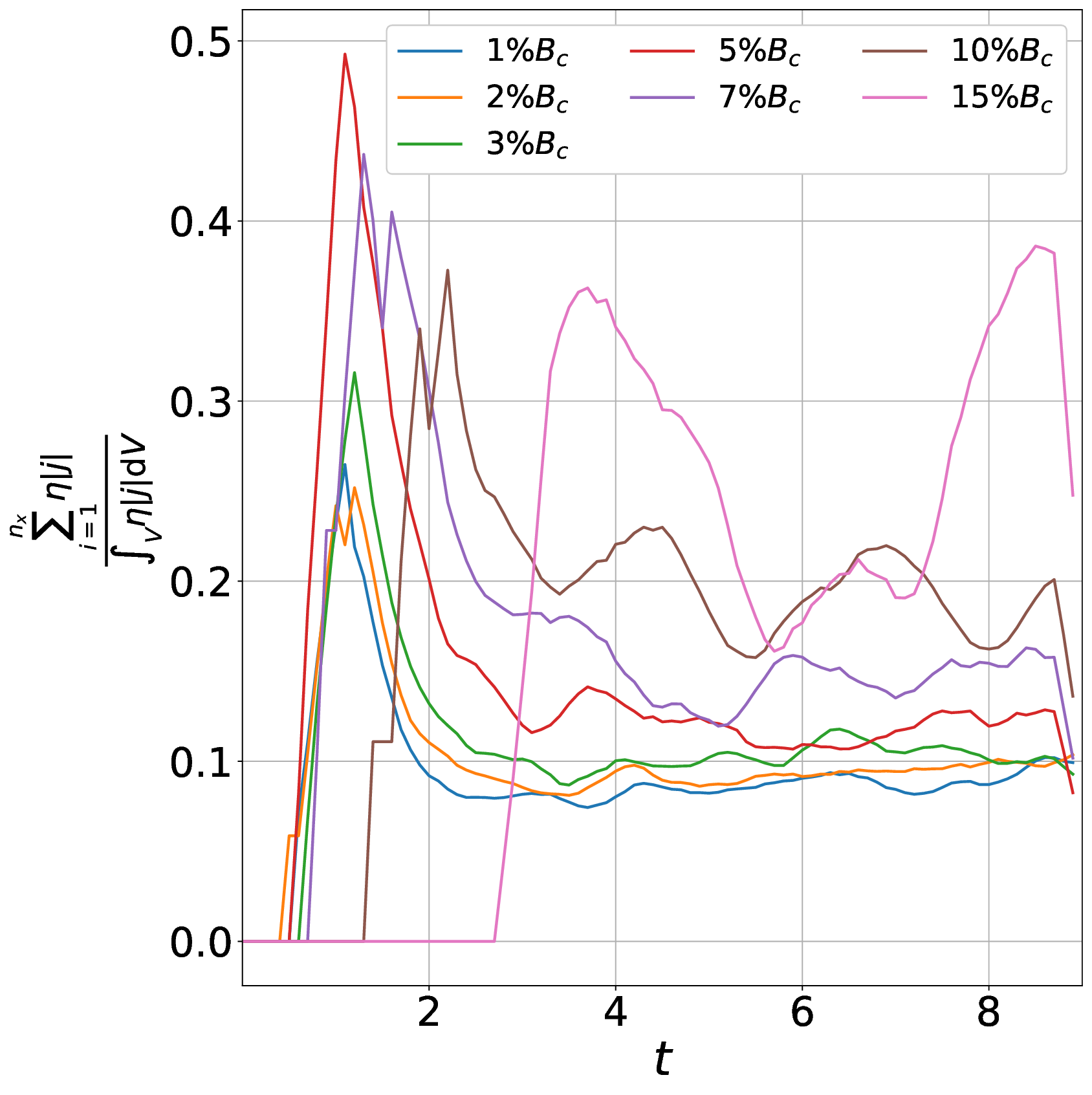}
    \caption{Temporal variation of total reconnection rate (non-dimensionalized by $\int_V \eta |j| \mathrm{d}V$), $\left( \frac{1}{n_{\times}} \sum_{i = 1}^{n_{\times}} \eta |j_i| \right)$ for different magnetic field strengths. The data is smoothed by performing a rolling average with a window of $0.3$ time units.}
    \label{rr_t}
\end{figure}

Figure \ref{rr_t} shows the time evolution of $RR$ normalized by the total current dissipation in the domain, i.e., $\int_V \eta |j| \mathrm{d}V$, for various magnetic field strengths. Interestingly, the normalized $RR$ remains approximately constant in time, even though the magnetic Reynolds number that drives the field lines to reconnect increases with time. This temporal constancy appears across all magnetic field strengths.

We also observe that $RR$ increases systematically with magnetic field strength. In weak-field regimes, reconnection predominantly involves weak current sheets (cf. figure \ref{hist}), leading to smaller reconnection rate. As the field strengthens, more intense sheets form, and reconnection becomes correspondingly more vigorous.

\vspace{-10pt} \subsection{Section summary}

\vspace{-10pt} To summarise $\S$\ref{sec:energydynamicsquant}, we found that the energy extracted from gravitational potential energy (GPE) is largely converted into magnetic energy. The magnetic field, in turn, returns nearly half of this energy to the system as kinetic energy. Reconnection plays a central role in this energy transfer. In weak magnetic field regimes, reconnection accounts for approximately $80\%$ of the total magnetic-to-kinetic energy conversion. Although this fraction decreases with increasing field strength, reconnection remains a significant mediator of energy transfer across all regimes. In contrast, reconnection contributes only marginally to total magnetic energy dissipation ($D_{\text{TME}}$). This limited contribution arises in part because most reconnection events involve weak current sheets, where dissipation is minimal (cf. $D_{\text{TME}} = \eta j^2$). Moreover, the system preferentially channels magnetic energy into flow, rather than dissipating. Finally, we found that the magnetic field strength also affects the reconnection rate. As the field strength increases, the total reconnection rate grows. However, for any given field strength, the total reconnected flux summed over all reconnection points remains approximately constant in time.

\vspace{-10pt}
\section{Discussion} \label{sec:extension}

\subsection{Variation of number of reconnection points with Reynolds number} \label{nxexplanation}
\vspace{-5pt}
In $\S$\ref{sec:number}, we found that the number of reconnection points varies as the square root of the magnetic Reynolds number $Re_m$. Here we discuss potential explanations for this scaling. As mentioned in $\S$\ref{sec:number}, the number of reconnection points depends on the size of the mixing layer and the level of turbulence in the mixing layer. For a given turbulence intensity, as the mixing layer widens, the number of locations at which the field lines contact increases. Thus, the number of reconnection events increases with the mixing layer height ($h$), i.e., $n_1 \propto h$. On the other hand, for a fixed mixing layer height, as the turbulence in the mixing layer increases, the magnetic field lines contact frequently, thereby increasing the number of reconnection events. Thus, we expect the number of reconnection points to increase with the turbulence. The turbulence in the system is often quantified by the turbulent Alfven velocity, defined as $v_A {=} \sqrt{\int_V \frac{(b_x^2 + b_z^2)}{\rho} dV}$. Thus, $n_2 \propto v_A$. Here, $n_1$ and $n_2$ are the number of reconnection points at a fixed turbulence intensity and height, respectively.

To understand our assumption of linear relation between $n_1$ and $h$, consider a domain of size $L \times h$ that is fully turbulent system with turbulence intensity $v_A$. Let $n_1$ be the number of reconnection points in this system. Now consider another turbulent domain with the same domain size and turbulence intensity with $n_1$ reconnection points. Stacking these two systems and ensuring the total turbulence intensity is same as individual domain, we have a total domain of $L \times 2h$, and the total number of reconnection points would be $2 n_1$. Thus, $n_1$ is linearly proportional to $h$. To understand the linear relation between $n_2$ and $v_A$, consider another case where we have a fully turbulent system in the domain $L \times h$. As the turbulence intensity in the domain increases from $v_A$ to $2 v_A$, the frequency of magnetic field line interaction becomes twice, and hence the number of reconnection points also increases by 2-fold. Thus, the number of reconnection points is linearly proportional to $v_A$.

In the MRTI, as the instability evolves, the mixing layer height and the turbulence increase simultaneously, which is reflected in the increase of magnetic Reynolds number ($Re_m {=} v_A h/\eta$). The temporal increase of $Re_m$ can be seen from figure \ref{nx_Re_t}\textit{(center)}. The number of reconnection points in the MRTI mixing layer is given by the geometric mean of the $n_1, n_2$, which gives,
\begin{equation}
    n_{\times} \equiv \sqrt{n_1 n_2} \propto \sqrt{v_A h} \propto \sqrt{Re_m}.
    \label{nx_relation}
\end{equation}
The latter comes from constant $\eta$. Thus, the number of reconnection points ($n_{\times}$) is expected to scale with magnetic Reynolds number as $\sqrt{Re_m}$. This was found to be consistent from our numerical simulations, as shown in figure \ref{nx_Re_t}\textit{(right)}. 

\citet{Kalluri_2024} showed that $\int_V \frac{1}{2}b^2 \mathrm{d}V \equiv L_x h (\frac{1}{2}b^2) \propto h^2$, i.e., $v_A {\propto} \sqrt{h}$. Substituting this relation in equation \ref{nx_relation}, we get $n_{\times} {\propto} h^{3/4}$. From the quadratic growth of mixing layer height \citep{Kalluri_2024}, $n_{\times} \propto t^{3/2}$. These scaling laws were found to be satisfied for all magnetic field strengths in our simulations. The $t^{3/2}$ growth demonstrates that the number of reconnection events grows in a self-similar fashion in the non-linear MRTI. This signifies that the self-similarity of MRTI is not restricted to flow variables like $\rho u,$ $b,$ TKE, TME, \citep{Kalluri_2024} but extends to other parameters like $n_{\times}$, that depend on the flow variables.

Alternatively, a more theoretical proof for $h^{3/4}$ scaling can be obtained from the energy dynamics analysis. The rate of energy injection at the integral length scales ($k_0$) is given by
\begin{equation}
    E = \frac{\rho v^2}{\tau} = \rho v^3 k_0.
\end{equation}
The latter comes from the scaling $\tau = \frac{1}{k_0 v}$. From Kolmogorov's theory, the injected energy trickles down to dissipative length scales. The scale at which the energy dissipates depends on the energy injected and the (magnetic) diffusion coefficient, in the case of reconnection current sheets. 

The scale at which energy dissipates predominantly can be written as 
\begin{equation}
    k_{\eta} = E^{\alpha} \rho^{\beta} \eta^{\gamma} \equiv (\rho v^3 k_0)^{\alpha} \rho^{\beta} \eta^{\gamma}
\end{equation}
From the dimensional analysis, 
\begin{equation}
    L^{-1} = (\rho L^2 \tau^{-3})^{\alpha} \rho^{\beta} (L^2 \tau^{-1})^{\gamma}
\end{equation}
\begin{equation}
    \alpha + \beta = 0; 2 \alpha + 2 \gamma = -1; -3 \alpha - \gamma = 0 .
\end{equation}
Solving for $\alpha$, $\beta$, and $\gamma$ we get $\alpha = 1/4$, $\beta = -1/4$ and $\gamma = -3/4$. Thus 
\begin{equation}
    k_{\eta} = E^{1/4} \rho^{-1/4} \eta^{-3/4}.
\end{equation}
Substituting the energy injected rate in the above equation, we get
\begin{equation}
    k_{\eta} = k_0^{1/4} v^{3/4} \eta^{-3/4}.
\end{equation}

\begin{figure}
    \centering
    \includegraphics[width=0.85\linewidth, height=8cm]{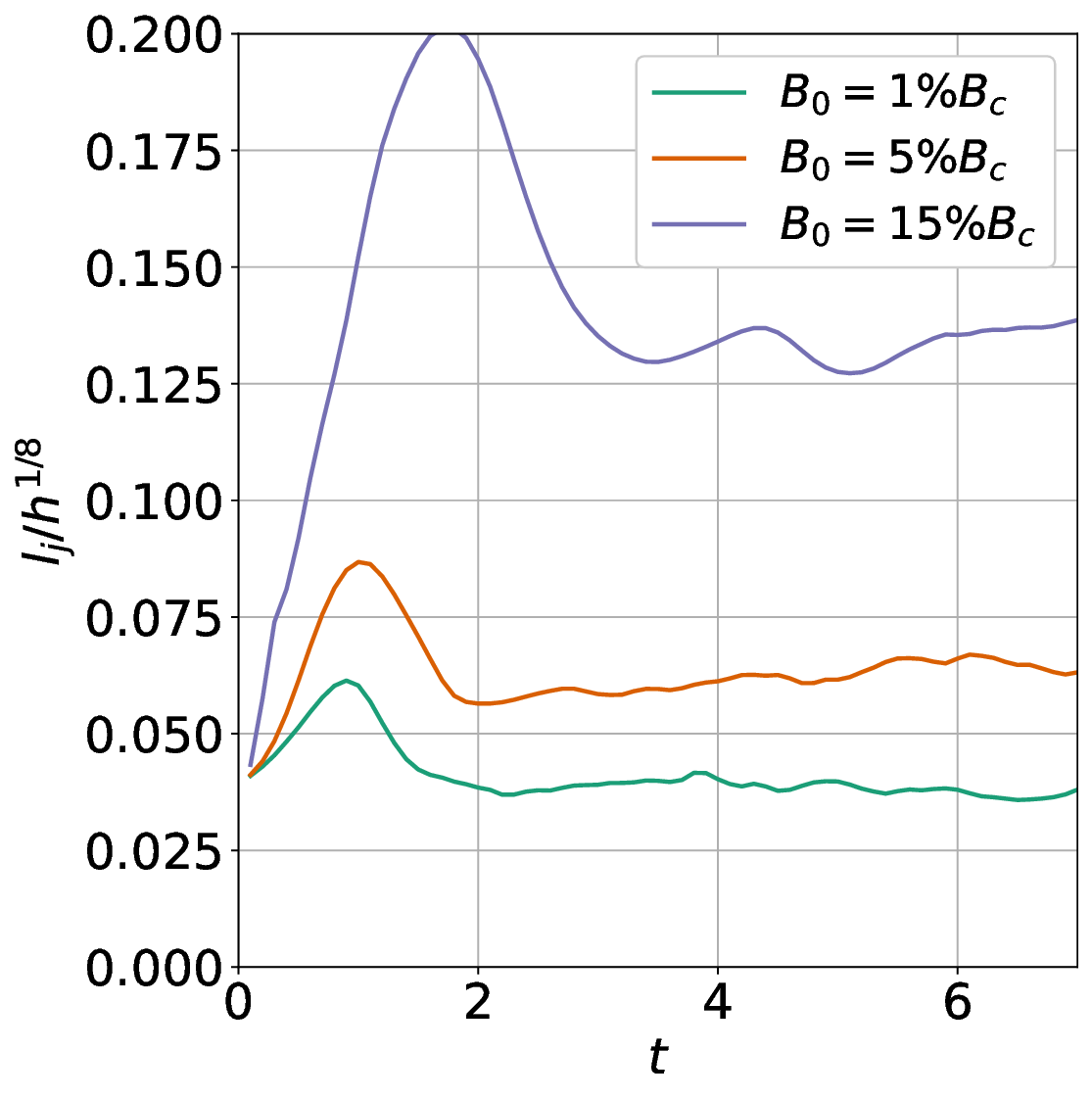}
    \caption{Temporal variation of characteristic current sheet length scale normalized by $h^{1/8}$ for different magnetic field strengths.}
    \label{ljhscaling}
\end{figure}

To write $n$ in terms of $h$, we write different terms of $k_{\eta}$ in terms of $h$. The energy injection occurs at integral length scales; therefore, $k_0$ is of order, $1/h$. From the quadratic growth of mixing layer height with time, $v \propto \sqrt{h}$. With these substitutions, we get $k_{\eta} = h^{1/8}$. Our numerical simulations indeed show that the characteristic length scale of the current sheets is of the order $h^{1/8}$. In figure \ref{ljhscaling}, we evidence this for three magnetic field strength cases.
The number of current sheets of scale $k_{\eta}$ in the mixing layer region ($L \times h$) is given by
\begin{equation}
    n \propto \frac{L h}{k_{\eta} k_{\eta}} \propto L h^{3/4}.
\end{equation}

\subsection{Dynamics of reconnection jets}
\begin{figure}[t]
    \centering
    \includegraphics[width=0.75\linewidth, height=9cm]{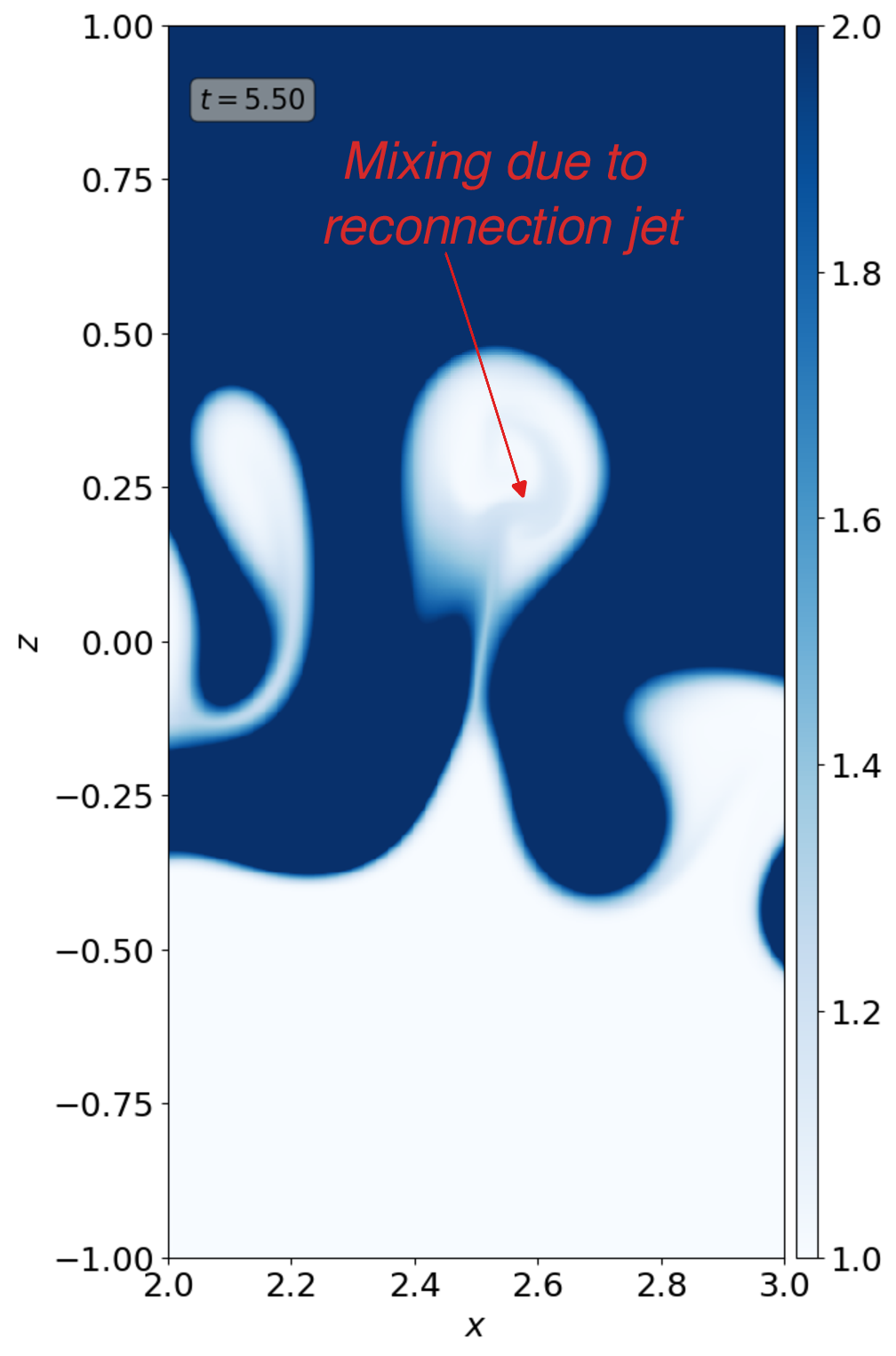}
    \caption{Figure demonstrating the mixing due to reconnection jet. Note that the fluid densities used in the simulation are 3, 1, but we limit the densities in the contour to [1, 2] to illustrate the mixing more clearly. }
    \label{mix}
\end{figure}

\vspace{-5pt}
The outbursts from reconnection are often interesting from a turbulence perspective, particularly in the strong magnetic field case, where these are more prominent. We will discuss the dynamics of these structures. From the fluid mixing perspective, we find that the reconnection jets lead to fluid mixing by transporting the diffused matter at the reconnection region into one of the fluids, as shown in the figure \ref{mix} where diffuse matter is transported to a lower density fluid. However, a quantitative estimate of the mixing solely due to reconnection is difficult. This is because, in the weak magnetic field case, most fluid is mixed by the turbulence mixing of fluids, making the mixing due to reconnection indistinguishable. In the strong magnetic field case, the intermediate density fluid is predominantly formed due to mass diffusion, and the role of reconnection in fluid mixing seems relatively less prominent.

The other aspect is the dynamics of these out-flowing jets themselves. While the outflow jets are known to exhibit shear instabilities \citep{Hillier2021}, we do not see such secondary KHI in our current simulations. This could be due to the suppression of shear instabilities by the magnetic field \citep{Chandrasekhar_1961}. The plume in Figure 20 (right) exhibits little turbulence due to the suppression of small scales by the strong magnetic field. Hence occurrence of secondary instabilities in such conditions is quite difficult. The other possible reason for not seeing secondary instabilities could be due to the inadequate grids to resolve the secondary instabilities. For example, the jets in Figure 20 are resolved by $100 \times 10$ grid points approximately.

\subsection{Relevance to 3D MRTI}
While the current work explores the role of reconnection in a 2D MRTI system (at fixed magnetic diffusivity, $\eta$), where the reconnection happens in-plane. However, in a 3D system, the reconnection is free to occur in any plane, and reconnection occurs under complex configurations. Hence, one should be wary of extending the results to a 3D MRTI case. We present our hypothesis on the validity of the results from the current work for the 3D MRTI case. In the current study, we found that the number of reconnection points scales according to $\frac{n_{\times}}{\sqrt{Re_m}} = \frac{0.15 B_c}{B_0}$. We hypothesize that the scaling between the number of reconnection events and the magnetic Reynolds number $(n_{\times} \propto \sqrt{Re_m})$ may hold in 3D, since there are still only three dependent parameters ($h, v_A, \eta$) in 3D. Also from theoretical scaling perspective in 3D, $k_{\eta}$ would still have $h^{1/8}$ scaling and the only difference is instead of $L$ we would have $L_x \times L_y$.

While the scaling remains same, the number of reconnection events could increase in 3D, owing to the greater degrees of freedom for the reconnection to occur in 3D. Hence, the proportionality constant $\frac{0.15 B_c}{B_0}$ could change in the 3D case. We expect the qualitative role of reconnection discussed in the long-term evolution of instability ($\S$\ref{section:pinching}) and energy dynamics, conversion from magnetic to kinetic and thermal energies ($\S$\ref{sec:energy_dynamics}) to be relevant for the 3D MRTI. However, the quantitative significance, discussed in $\S$\ref{sec:energydynamicsquant}  might vary in 3D.

\subsection{Relevance to astrophysical systems}
The study presented here is founded on several simplifications, like the 2D, incompressibility, fully ionized charge-neutral plasma, uniform unidirectional magnetic field, and relatively small kinetic and magnetic Reynolds numbers. However, several features present in the simulation are seen in observations of astrophysical systems. For example, Figure 14 shows the formation of isolated blobs of low-density fluid in the high-density fluid due to reconnection, see the blobs at $(x, y) \approx (2.75, 0.25)$ in Figures 14(b), 14(c). Such isolated blobs were observed in the Rayleigh Taylor instability of quiescent solar prominences (for example, figure 8 from \cite{Berger2010}), which could be potentially explained due to the reconnection. The formation of reconnection outflow jets was demonstrated in the RTI in solar prominences \cite{Hillier2021}, which were also observed in the current simulations.

\vspace{-10pt}
\section{Conclusions} \label{sec:conclusion}

\vspace{-10pt} 
Magnetic Rayleigh–Taylor instability (MRTI) and magnetic reconnection are ubiquitous in numerous astrophysical and laboratory systems, but the influence of one on the other has remained unclear. The current study brings these two fundamental aspects together and provides a comprehensive understanding of the role of magnetic reconnection in the MRTI through detailed isolated and statistical analyses. 
The approximate flux-frozen nature of plasma implies that the onset of MRTI demands deformation of not only fluid lines but also the magnetic field lines. Therefore, the non-linear interaction of plumes leads to magnetic reconnection, making reconnection a fundamental and inevitable phenomenon of MRTI. Despite this, the influence of reconnection on MRTI has remained poorly understood. Here, we summarize the key questions addressed in this work and their corresponding answers: \newline

1. \hspace{3pt} Is magnetic reconnection essential for the long-term instability evolution? \newline
Yes, we establish that the magnetic reconnection is essential for the long-term evolution of instability. In an ideal-MRTI, where reconnection does not occur, as the instability evolves, the magnetic field lines elongate and the magnetic tension grows, eventually become strong enough to counter the gravitational force, the driving force of instability, arresting further growth of instability. In contrast, in the non-ideal case, reconnection changes the magnetic field topology, shortens field lines, and reduces magnetic tension. Reconnection facilitates plume merger by allowing plasma to cross field lines, resulting in larger plumes that experience stronger buoyancy forces. By moderating magnetic tension and promoting plume coalescence, reconnection is crucial for sustaining the instability’s nonlinear growth over long timescales.

2. \hspace{3pt} Does reconnection play a crucial role in the energy dynamics of MRTI? \newline
Reconnection plays a critical role in the energy dynamics. The majority of the magnetic to kinetic energy transfer is due to reconnection, especially in weak-field regimes, where it accounts for up to $\approx 80\%$ of this transfer. Through energy injection, magnetic reconnection is expected to improve fluid mixing locally. However, reconnection contributes only modestly ($\approx 3\%$) to total magnetic energy dissipation. The majority of the energy dissipation occurs from the non-reconnecting current sheets that form around the plume head and flanks. The reconnection is also a key element in regulating the magnetic tension in the system (cf. figure \ref{TKE_TME}) and ensuring the instability evolves continuously.  \newline

3. \hspace{3pt} How do the reconnection events vary statistically in the MRTI mixing layer? \newline
The MRTI mixing layer exhibits increasingly complex turbulent dynamics, which drive more frequent interactions and reconnection of magnetic field lines over time. It was found that, for any given magnetic field strength, the number of reconnection events increases self-similarly with time ($n_{\times} \propto t^{3/2}$) and magnetic Reynolds number ($n_{\times} \propto Re_m^{1/2}$). But, as the magnetic field strength is increased, the number of reconnection events for a given $Re_m$ decreases. This is due to the suppression of turbulence in the mixing layer by the magnetic field.

In summary, reconnection emerges as a fundamental mechanism that controls large-scale instability evolution by alleviating magnetic tension and enabling plume mergers. While the number of reconnection events grows self-similarly, the proportion of reconnecting weak current sheets is increasing. Consequently, the dynamical significance of reconnection in energy dissipation is not significant. However, the reconnection plays a crucial role in the transfer of magnetic energy into the system. Thus, the impact of reconnection on MRTI evolution and dynamics is profound.

\vspace{-10pt}
\begin{acknowledgments}
\vspace{-10pt} MTK is supported by the EPSRC Grant No. EP/W523859/1. AH is supported by STFC Research Grant No. ST/R000891/1 and ST/V000659/1. BS is supported by STFC Research Grant No. ST/V000659/1. The computational time was obtained from the University of Exeter High-Performance Computing facility. 
\end{acknowledgments}

\vspace{-15pt}
\section*{Data Availability Statement}
\vspace{-10pt} The data that support the findings of this study are available from the corresponding author upon reasonable request.

\vspace{-15pt}
\section*{Author contributions}
\vspace{-10pt} M.T.K.: reconnection detection algorithm, data curation, formal analysis, investigation, methodology, software, validation, visualization, writing —original draft, review and editing; A.H.: conceptualization, funding acquisition, project administration, resources, supervision, reconnection detection algorithm, writing—review and editing; B.S.: reconnection detection algorithm, writing—review and editing

\vspace{-10pt}
\appendix
\section{Issues in the reconnection detection algorithm and their solutions} \label{Detection}

\begin{figure*}
    \centering
    \includegraphics[width = 0.9\textwidth]{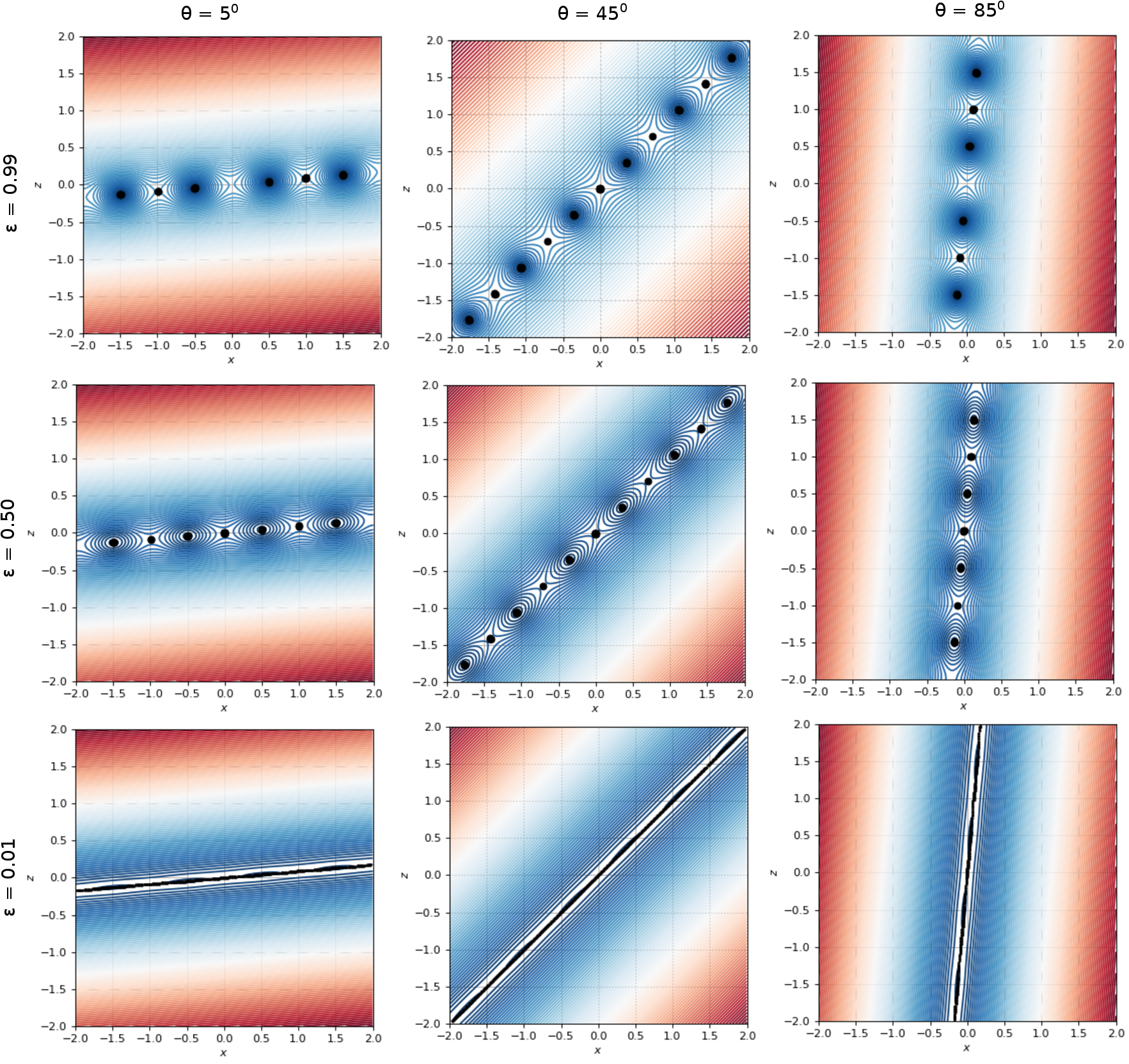}
    \caption{Result from null point detection code for Fadeev equilibrium at different orientations ($\theta = 5^{\circ}, 45^{\circ}, 85^{\circ}$) and eccentricities ($\epsilon = 0.01, 0.45, 0.99$).}
    \label{Fadeev}
\end{figure*}

\vspace{-10pt} The algorithm mentioned in $\S$\ref{section:detection} has a few issues which are described below, again using Fadeev equilibrium. The code is tested for Fadeev and Harris equilibrium in a $4 {\times} 4$ domain $(L_x {=} L_y {=} [-2.0, 2.0])$, with a $400 {\times} 400$ grid. The problem is tested for different eccentricity $(\epsilon)$ and angle of orientation $(\theta)$, but we only show a sample case here. Figure \ref{Fadeev} shows the null points detected by the code for different $\epsilon$ values and orientations of plasmoids. The value of $\kappa$ is $2 \pi$ for $\theta = 5^0,$ $85^0$, and $3 \pi$ for $\theta = 45^0$. 

\begin{figure*}
    \centering
    \includegraphics[width = 0.9\textwidth]{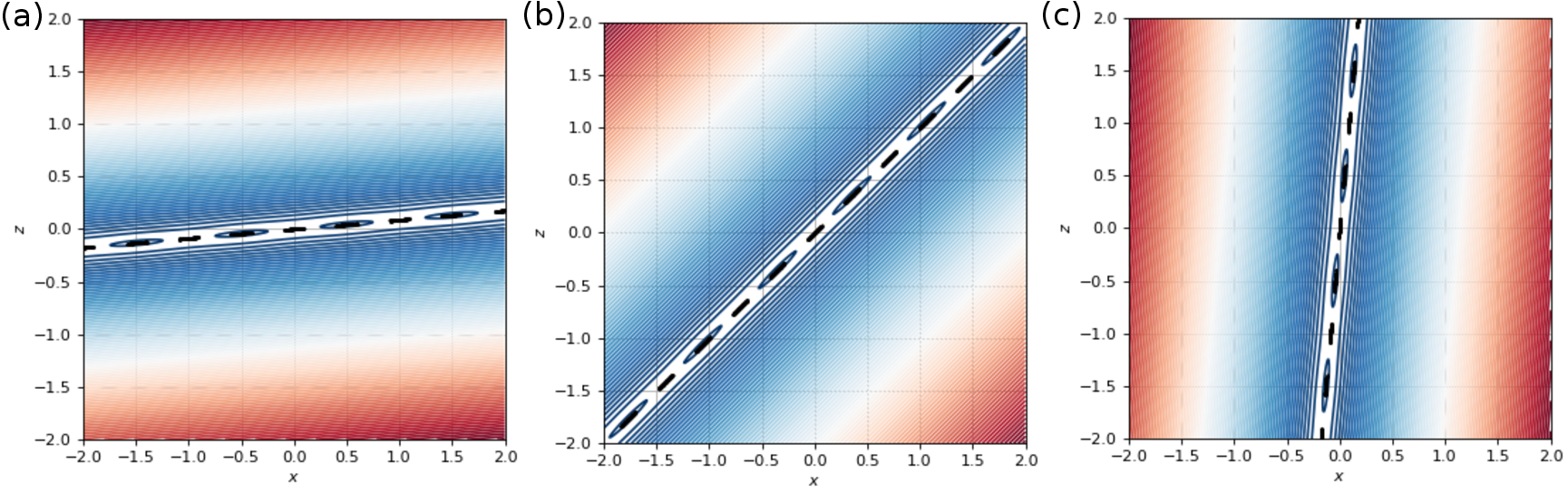}
    \caption{Contour plots of magnetic vector potential and null points detected by the code for elongated plasmoids ($\epsilon = 0.01$) for different angles $(\theta = 5^{\circ}, 45^{\circ}, 85^{\circ}$ from left to right). The resolution for the current case is $1600 \times 1600$.}
    \label{Fadeev_extreme}
\end{figure*}

\begin{figure*}
     \centering
     \begin{subfigure}[b]{0.45\textwidth}
         \centering
         \includegraphics[width=\textwidth]{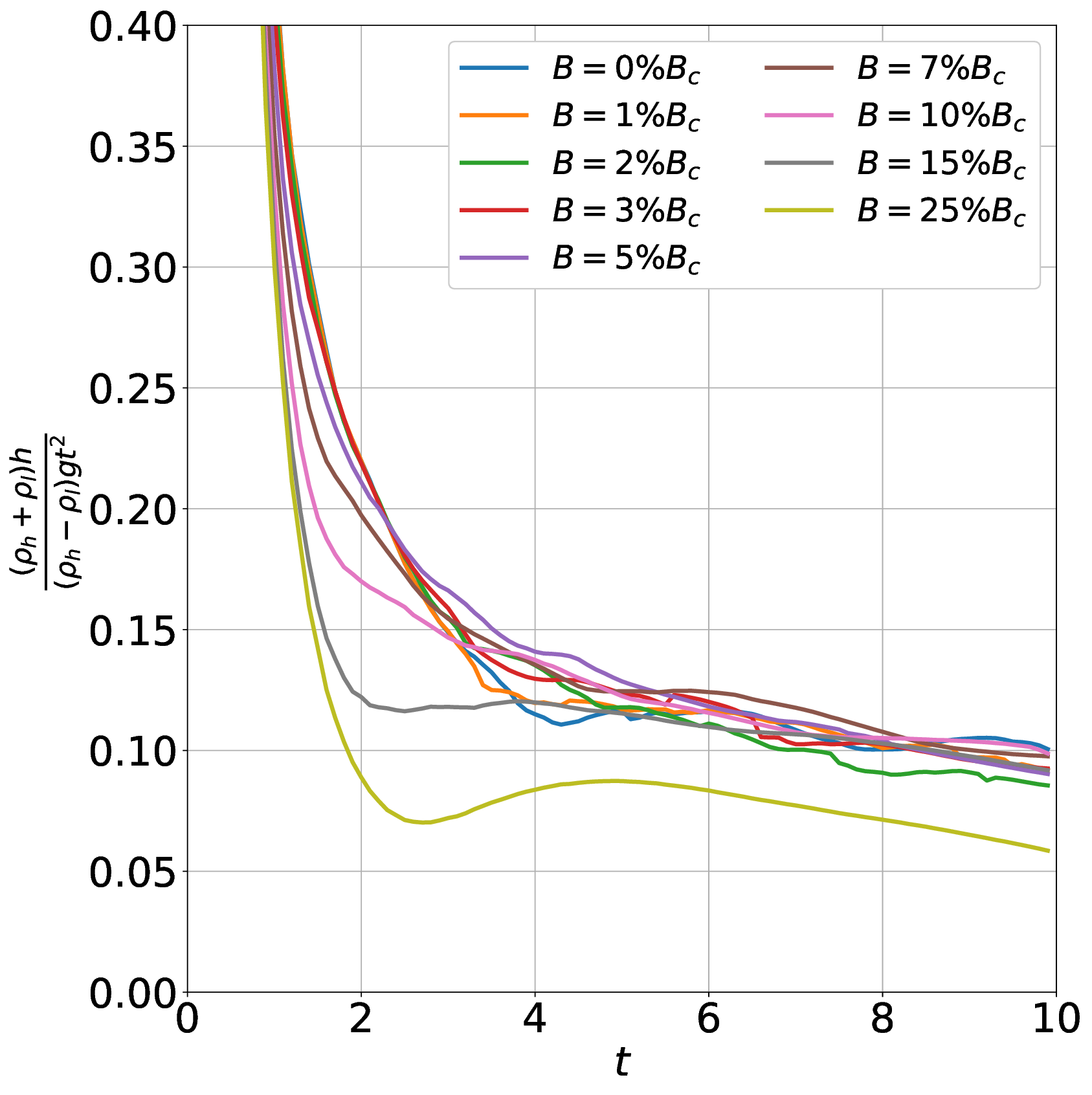}
     \end{subfigure}
     \begin{subfigure}[b]{0.45\textwidth}
         \centering
         \includegraphics[width=\textwidth]{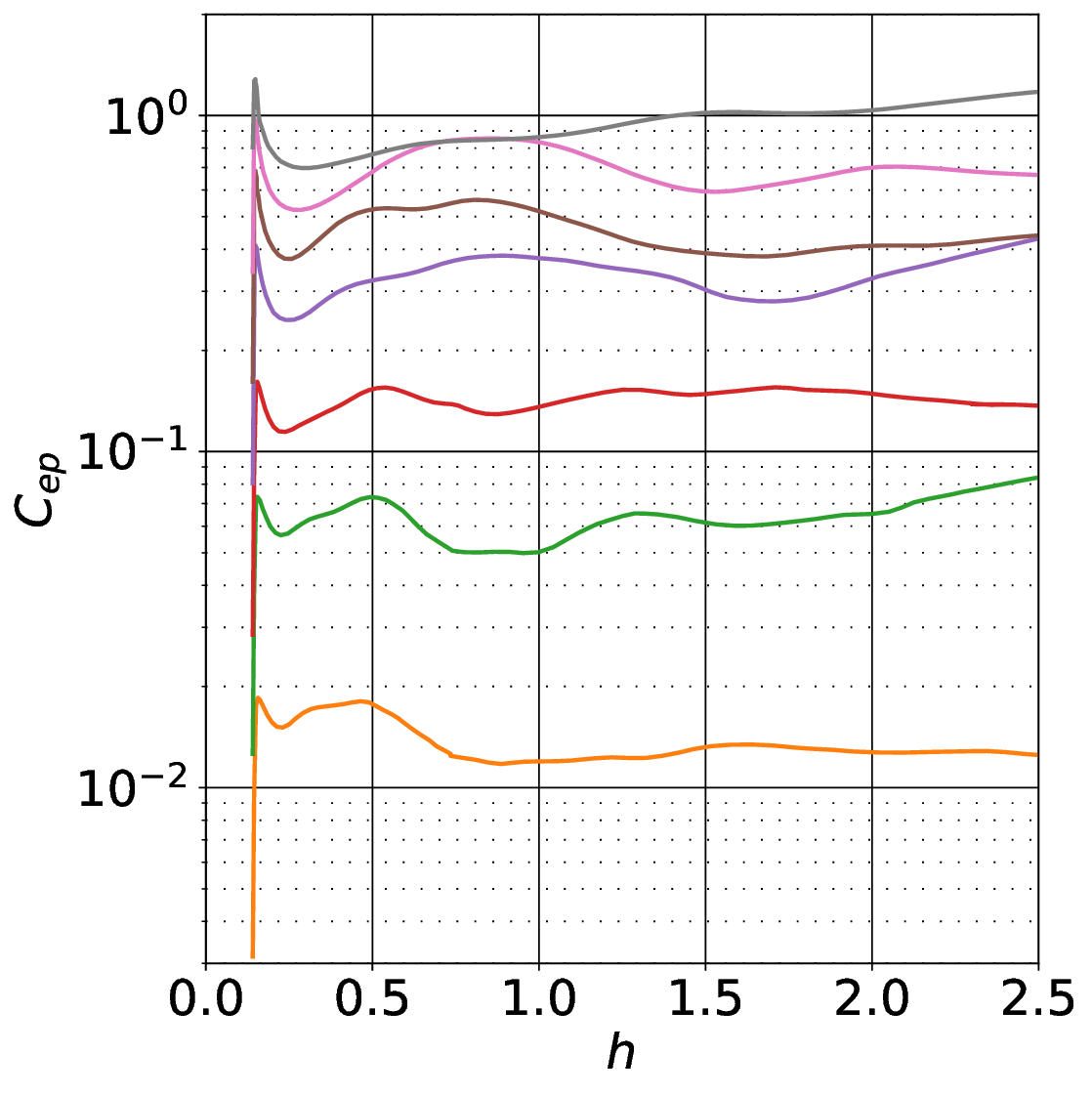}
     \end{subfigure}
 \caption{Temporal variation of: \textit{(left)} $\frac{(\rho_h + \rho_l)}{((\rho_h - \rho_l)g} \frac{h}{t^2}$; \textit{(right)} energy partition ($C_{ep}$) i.e., ratio of turbulent magnetic energy to turbulent kinetic energy.}
 \label{selfsimilarity}
\end{figure*}

From the last row of figure \ref{Fadeev}, at very low $\epsilon$, i.e., when plasmoids are highly elongated, the code gives a chain of points. This is due to the poor resolution of the plasmoids. When the number of grid points is increased, the code detects the location of these points more accurately, as shown in figure \ref{Fadeev_extreme}, where the grid points are increased to $1600 \times 1600$ for the same parameters. The need to resolve the plasmoids is a major issue for the algorithm. Often, in a highly turbulent system, the plasmoids are elongated due to strong shear. But, due to limited resolution capability, it is impossible to resolve all plasmoids, and we often end up with a chain of X-points, resulting in false reconnection points.

The reconnection region typically spans over multiple grid points, resulting in a cluster of points (see figure \ref{Fadeev_extreme}). However, all the points represent one null point (or reconnection point). To address this issue, we use a clustering method \cite{scikit-learn} to cluster these points into a single point. We use the density-based clustering techniques (DBSCAN \citep{Ester1996ADA, Schubert2017}, HDBSCAN \citep{Campello2015}) due to the collocation nature of the points. The DBSCAN method requires the user to provide an $eps$ value, which represents the radius of the neighbourhood region. Hence, DBSCAN is suitable for scenarios when the clusters are similar throughout the space. However, in turbulent systems, the reconnecting clusters are often of very different sizes and occur in various sizes. Thus, specifying a value for $eps$ is not possible. Therefore, we use HDBSCAN, where the algorithm determines the clusters by decreasing the $eps$ value. As the $eps$ value decreases, the cluster becomes smaller in size. The part that remains a cluster across the $eps$ value is considered the final cluster. A drawback of the above method is the potential grouping of two or more clusters close to each other into one. This issue is unavoidable due to the choice of autonomous clustering and could also happen in the DBSCAN method (depending on the $eps$ value provided).

Once the clusters are determined, the reconnecting point in each cluster is to be chosen. Two potential methods we can use to choose the point are --- $i)$ the centroid of the cluster, $ii)$ the point with the least magnetic field magnitude. For the former, the mean coordinates of each cluster are calculated \footnote{Since the mean need not always be at the grid point, we determine the nearest grid point close to the determined centroid and assign it as the reconnection point.}. A problem with this method is the inaccuracy of the $X-$point. If HDBSCAN considers two or more neighbouring clusters as one cluster, the X-point location based on the centroid is inaccurate. This can be avoided with the latter choice, and hence, we choose the least magnetic field magnitude method. While this method reduces, it cannot eliminate false reconnection points.

\section{Self-similarity in 2D MRTI}
In $\S$\ref{sec:number}, we mentioned that MRTI is self-similar. However, this was not evidenced there since the focus of the section was on reconnection. Here, we evidence the self-similarity of the present MRTI simulations. One of the standard methods to validate MRTI self-similarity is by verifying the quadratic growth of mixing layer height (cf. $h \propto \frac{(\rho_h - \rho_l)g t^2}{(\rho_h + \rho_l)}, t \gg 1$) \citep{Stone2007a, Kalluri_2024}. In the self-similar regime, we expect $\frac{(\rho_h + \rho_l)}{((\rho_h - \rho_l)g} \frac{h}{t^2}$ to be approximately constant. One of the other ways of confirming self-similarity is by plotting the energy partition ($C_{ep}$), which is the ratio of turbulent magnetic energy to turbulent kinetic energy over time \citep{Kalluri_2024}. From figure \ref{selfsimilarity}, we find that these parameters are approximately constant over time, evidencing that the system achieved a self-similar state. Sampling the data between $t = 6$ and $t = 10$, the largest standard deviation during the time frame is less than $1\%$.

\bibliography{apssamp}

\end{document}